\documentclass[namedreferences]{solarphysics}
\usepackage[optionalrh,natbib]{spr-sola-addons} 
  \usepackage[final]{graphicx}
\usepackage{amssymb}
\usepackage{color}           
\usepackage{url}             

\newcommand{\etal}{{\it et al.}}

\begin{document}

\begin{article}

\begin{opening}

\title{Time-Dependent Modulation of Cosmic Rays in the Heliosphere}

\author{R.~\surname{Manuel}$^{1}$\sep
        S.E.S.~\surname{Ferreira}$^{1}$\sep
        M.S.~\surname{Potgieter}$^{1}$      
       }
\runningauthor{R. Manuel \etal}
\runningtitle{Time-Dependent Modulation of Cosmic Rays}

   \institute{$^{1}$ Centre for Space Research, \\
North-West University, \\
Potchefstroom-2520, \\
South Africa.
                     email: \url{rexmanuel@live.com} 
             }

\begin{abstract}
The time-dependent modulation of galactic cosmic rays in the heliosphere is studied by computing intensities using a time-dependent modulation model. By introducing recent theoretical advances in the transport coefficients in the model, computed intensities are compared with \textit{Voyager 1}, \textit{International Monitoring Platform (IMP) 8}, and \textit{Ulysses} proton observations in search of compatibility. The effect of different modulation parameters on computed intensities is also illustrated. It is shown that this approach produces, on a global scale, realistic cosmic-ray proton intensities along the \textit{Voyager 1} spacecraft trajectory and at Earth upto $\sim$2004, whereafter the computed intensities recovers much slower towards solar minimum than observed in the inner heliosphere.  A modified time dependence in the diffusion coefficients is proposed to improve compatibility with the observations at Earth after $\sim$2004. This modified time dependence led to an improved compatibility between computed intensities and the observations along the \textit{Voyager 1} trajectory and at Earth even after $\sim$2004. An interesting result is that the cosmic-ray modulation during the present polarity cycle is not  determined only by changes in the drift coefficient and tilt angle of the wavy current sheet, but is also largely dependent on changes in the diffusion coefficients.
\end{abstract}
\end{opening}

\section{Introduction}
Cosmic rays in the heliosphere experience changes in their intensities as a function of energy, position, and time due to the out-blowing solar wind and the embedded heliospheric magnetic field (HMF). This process is known as cosmic-ray modulation. When cosmic rays enter the heliosphere, they experience four major modulation processes: (i) convection, due to the expanding solar wind, (ii) energy changes such as adiabatic energy losses, continuous stochastic acceleration and even diffusive shock acceleration,  (iii) diffusion due to the random walk along and across the turbulent HMF, and (iv) drifts, due to the gradient and curvatures in the HMF or any abrupt changes in the field direction such as the current sheet. Cosmic-ray propagation is influenced by solar activity and this leads to 11- and 22-year modulation cycles in the cosmic-ray intensities. See \textit{e.g.} \cite{Jokipii-etal-1977,Potgieter-and-Moraal-1985,Shalchi-etal-2004a,Zhang-2006,Strauss-etal-2010a,Potgieter-2013a,Potgieter-2013b}. 

Concerning the theoretical description of long-term cosmic-ray modulation, it was shown by \citet{Perko-and-Fisk-1983} and \citet{Leroux-and-Potgieter-1989} that in order to describe cosmic-ray modulation over long periods, some form of propagating diffusion barriers is required. This is especially true for solar maximum activity periods when step decreases in cosmic-ray intensities are observed. The largest of these diffusion barriers are called global merged interaction regions (GMIRs): \citep{Burlaga-etal-1993}. Equally important to modulation are gradient, curvature, and current sheet drifts \citep{Jokipii-etal-1977,Potgieter-and-Moraal-1985} as confirmed by the comprehensive modelling done by \citet{Potgieter-etal-1993} and \citet{Leroux-and-Potgieter-1995}. These authors showed that it is possible to simulate, to first-order, a complete 22-year modulation cycle by including a combination of drifts and GMIRs in a time-dependent modulation model. The conclusion was made that drifts are mainly responsible for time-dependent modulation during moderate to minimum solar conditions with cosmic-ray intensities changing mainly due to changes in the current sheet during $A < 0$ polarity cycles. However, towards solar activity maximum, GMIRs cause the intensities to decrease in a step-like manner with drifts becoming less dominant. For reviews see \cite{Potgieter-1993,Potgieter-1997}. 

However, \citet{Cane-etal-1999} and \citet{Wibberenz-etal-2002} suggested that time-dependent global changes in the HMF magnitude alone might be responsible for long-term modulation. This was tested by \cite{Ferreira-and-Potgieter-2004}, who introduced the compound approach, where all the transport (diffusion and drift) parameters are scaled with a time-dependent function based on the observed solar magnetic field at Earth and the current sheet tilt angle \citep{Hoeksema-1992}. This resulted in diffusion coefficients and drifts changing over a solar cycle, with smaller values for solar maxima compared to solar minima. As shown by \citet{Ferreira-and-Potgieter-2004}, \citet{Ndiitwani-etal-2005}, \citet{Ferreira-and-Scherer-2006}, and \citet{Manuel-2011a} this compound approach incorporated in a numerical modulation model yields results in good agreement with spacecraft observations (\textit{Ulysses} in particular) at various energies. 

The compound model of \citet{Ferreira-and-Potgieter-2004} is based on an empirical approach where computed results are compared to observations in order to construct a realistic time dependence in the transport coefficients. This was done because of a lack of a clear theory on how diffusion and drift coefficients should change over a solar cycle. However, recent progress by \citet{Teufel-and-Schlickeiser-2002, Teufel-and-Schlickeiser-2003}, \citet{Shalchi-etal-2004a}, and \citet{Minnie-etal-2007b} gives a much clearer picture of how the diffusion coefficients depend on basic turbulence quantities, such as the magnetic field magnitude and variance, that change over a solar cycle.

These new theories, which have never been tested time-dependently, are introduced in a numerical model to calculate long-term, time-dependent cosmic-ray modulation \citep[see also][]{Manuel-2011a, Manuel-2011b}. We have shown that in order to construct a time dependence in the transport coefficients, time-dependent changes in basic turbulence quantities such as the magnetic field magnitude and variance as well as the tilt angle of the current sheet are needed. These quantities are then transported out into the simulated heliosphere  and result in realistic computed modulation on a global scale when compared to observations. 

In this work we will again show that after incorporating recent theoretical advances, the model produces results that are compatible with observations along the \textit{Voyager 1} spacecraft trajectory and at Earth up to $\sim$2004 \citep[See also][]{Manuel-2011a,Manuel-2011b}. However, not discussed before is the finding that the model fails to reproduce observations at Earth from $\sim$2004 onwards.  It will be shown in detail that after varying most of the important model parameters, the model still could not reproduce the observations at Earth for the period $\sim$2004 onwards using the assumed time dependence in the transport parameters as given by the recent theoretical advances. It will be shown that in order to fit the observations at Earth from $\sim$2004 onwards, a different time dependence in the diffusion coefficients is needed and that cosmic-ray modulation in this particular solar cycle is largely dependent on changes in value and rigidity of the diffusion coefficients and not primarily on changing the drift coefficient or the tilt angle of the current sheet \citep[see also][]{Potgieter-etal-2013}.

\section{Modulation Model}

The above-mentioned modulation processes were combined by \cite{Parker-1965} into a transport equation (TPE) given as
\begin{eqnarray}\label{tpe}
 \frac{\partial f}{\partial t}=
-\left( \mathbf{V} + \left\langle \mathbf{v}_{d} \right\rangle \right)  \cdot\nabla f +\nabla\cdot(\mathbf{K_{S}} \cdot\nabla
 f)  + \frac{1}{3}(\nabla\cdot \mathbf{V})\frac{\partial
   f}{\partial \ln P} + Q. 
\end{eqnarray}
Here, $t$ is the time, $P$ is rigidity, $Q$ is any particle source inside the heliosphere, $\mathbf{V}$ is the solar wind velocity, $\mathbf{K_{S}}$ is the isotropic diffusion tensor, and $\left\langle \mathbf{v}_{d} \right\rangle$ the pitch angle averaged guiding center drift velocity for a near isotropic distribution function [$f$]. The differential intensity [$j$] is related to $f$ by $j=P^{2}f$. 

This equation is solved numerically in this work in terms of time and rigidity in two-dimensional space $(r, \theta)$ with $r$ the radial distance and $\theta$ the polar angle. The rigidity [$P$] is defined as the momentum per charge for a given particle \textit{i.e.} $P = pc/q$ with $p$ the particle's momentum, $q$ the charge, and $c$ the speed of light. We assume the grid size in $r$ as $\Delta r \approx 0.7-0.9$ AU, which depends on the boundary position, and in $\theta$ as $\Delta \theta =2.43^{\circ}$. The rigidity step, $\Delta \ln P=0.08$ and the time step are chosen such that solar cycle related changes propagate with the solar wind speed.

Two diffusion coefficients of particular concern for this study are the effective radial diffusion coefficient [$K_{rr}$] and the effective perpendicular diffusion coefficient [$K_{\theta \theta}$] and are given by
\begin{eqnarray}\label{K_rreqn}
K_{rr} &=& K_{||} \cos^2 \psi+  K_{\perp r} \sin^2 \psi, \\
K_{\theta \theta} &=& K_{\perp \theta}, \nonumber
\end{eqnarray}
where $K_{||}$ is the diffusion coefficient parallel to the average HMF, $K_{\perp \theta}$ the diffusion coefficient perpendicular to the average HMF in the $\theta$ (polar) direction, $K_{\perp r}$ the diffusion coefficient perpendicular to the average HMF in the $r$ (radial) direction, and $\psi$ the spiral angle of the average HMF with the radial direction. 

The drift coefficient [$K_{A}$] used in this work is from \cite{Burger-etal-2000} and is given by
\begin{equation} \label{eq:KABurger}
K_A = K_{A0}\frac{\beta P}{3B}\frac{10 P^2}{10 P^2+1}
\end{equation}
where $\beta$ is the ratio of the particle speed [$v$] to the speed of light [$c$],  $K_{A0}$ is a dimensionless constant that could scale from 0 to 1, representing zero drift to full drift \citep{Potgieter-and-Leroux-1989}.


\section{Recent Theoretical Advances in the Transport Coefficients}
Of primary importance to cosmic-ray modulation is the coupling of the transport parameters to the background magnetic field and turbulence. For the parallel mean free path, \citet{Teufel-and-Schlickeiser-2002} gave an applicable expression for protons (damping model) in the inner heliosphere with rigidity [$P$] in the range $10^{-1} $ MV $<P<10^4 $ MV as $ \lambda_{||}\propto P^{1/3}$ at Earth. However, in the numerical model that calculates cosmic-ray intensities throughout the whole heliosphere, it is assumed that
\begin{equation}\label{par}
\lambda_{||} = C_1 \left( \frac{P}{P_0} \right)^{\frac{1}{3}} \left( \frac{r}{r_0} \right)^{C_2} f_{2}(t) \quad \textrm{  for }  \;  r < r_{\mathrm{ts}} 
\end{equation}
and
\begin{equation}\label{par-gt-rts}
 \lambda_{||} = \frac{C_1}{s_k}\left(\frac{P}{P_0}\right)^{\frac{1}{3}}\left(\frac{r}{r_0}\right)^{C_2} \left(\frac{r_{\mathrm{ts}}}{r} \right)f_{2}(t)\; \; \; \textrm{  for }  \;  r \geq r_{\mathrm{ts}}
\end{equation}
where $P_{0} =$ 1 MV, $r_{0}=1$ AU, $C_{1}$ [in units of AU] is a constant determining the absolute value of the mean free path, $C_{2}$ is a constant determining the radial dependence, $r_{\mathrm{ts}}$ is the termination shock [TS] position in AU, $s_k$ is the compression ratio, and $f_{2}(t)$ [as given below in Equation (\ref{f2})] is a dimensionless time-varying function that gives the time dependence in $\lambda_{||}$. This is transported from the Earth into the heliosphere and out into the heliosheath with the solar wind speed.

According to \citet{Burlaga-etal-2007}, the Voyager observations of $B$ indicate that $B \propto r$ for $r>r_{\mathrm{ts}}$. If the diffusion coefficients have some dependence on $B$, changes over the shock are expected. Equation (\ref{par}) is assumed valid for $r < r_{\mathrm{ts}}$, with the diffusion (mean free path) decreasing as the compression ratio [$s_k$] at the shock and then scaling as $\propto 1/r$  up to the heliopause [$r_{\mathrm{hp}}$], as given by Equation (\ref{par-gt-rts}). To calculate the cosmic-ray intensities \cite{Florinski-etal-2003}, \cite{Ferreira-and-Scherer-2006}, and \cite{Ferreira-etal-2007a, Ferreira-etal-2007b} made similar assumptions about the diffusion coefficients, assuming that they are to the first-order inversely proportional to $B$. Due to the flow deceleration of the solar wind, $B$ increases further towards the heliopause after a sudden initial increase over the TS. Note that the re-acceleration of galactic cosmic rays at the solar wind TS is not considered in this approach.

The time dependence [$f_{2}(t)$ in Equation (\ref{par})] of $K_{||}=\lambda_{||}v/3$ is attained using an intricate expression for $\lambda_{||}$ given by \citet{Teufel-and-Schlickeiser-2003} as
\begin{eqnarray}\label{E1}
\lambda_{||} = \frac{3s}{\sqrt{\pi}(s-1)} \frac{R^2}{b_{k} \ k_{\mathrm{min}}}\left(\frac{B}{\delta{B_{\mathrm{slab},x}}}\right)^2   \left[\frac{b_{k}}{4\sqrt{\pi}}+\frac{2}{\sqrt{\pi}(2-s)(4-s)}\frac{b_{k}}{R^s}\right]  
\end{eqnarray}
where $s=5/3$,  the spectral index of the inertial range, $b_{k}$ a fraction of particle to Alfven speed assuming maximum dynamical effects, $k_{\mathrm{min}} = 10^{-10}$ m$^{-1}$ the spectral break point between the inertial and energy range on the turbulence power spectrum at 1 AU, $ \ R = k_{\mathrm{min}}R_{\mathrm{L}}$,  with the Larmor radius $\ R_{\mathrm{L}} = \frac{P}{Bc}$ and $\delta{B^{2}_{\mathrm{slab},x}}$ the $x$-component of the slab variance. The slab variance $\delta{B^{2}_{\mathrm{slab}}} = 2 \delta{B^{2}_{\mathrm{slab},x}}$ and when \textit{e.g.} a 20/80 ratio of slab to 2D variance is assumed \citep[][]{Bieber-etal-1994}, $\delta{B^{2}_{\mathrm{slab}}}=0.2\delta{B^{2}}$ and $\delta{B^{2}_{\mathrm{2D}}}=0.8\delta{B^{2}}$ where $\delta{B^{2}}$ is the total variance and $\delta{B^{2}_{\mathrm{2D}}}$ the two-dimensional variance. 

Only the influence of time-varying quantities $B$ and $\delta{B^{2}}$ on $\lambda_{||}$ is considered, and Equation (\ref{E1}) is approximated as
\begin{eqnarray}\label{parApprox}
\lambda_{||} \propto \left( \frac{1}{\delta{B_{\mathrm{slab},x}}}\right)^2\left[\frac{1}{4}+\frac{18}{7}\left(\frac{B \ c}{P \ k_{\mathrm{min}}}\right)^{5/3}\right].
\end{eqnarray}
From Equation (\ref{parApprox}), the function $f_{2}(t)$ [in Equation (\ref{par})] can be deduced as
\begin{eqnarray}\label{f2}
f_{2}(t)= C_{4}\left( \frac{1}{\delta{B}(t)}\right)^2\left[\frac{1}{4}+\frac{18}{7}\left(\frac{B(t) \ c}{P \ k_{\mathrm{min}}}\right)^{5/3}\right] ,
\end{eqnarray}
with $C_{4}$ a constant in units of $\textrm{(nT)}^2$.

For perpendicular diffusion, it was shown using simulations that $K_{\bot r}$ and $K_{\bot \theta}$ scale as the parallel coefficient  \citep{Leroux-etal-1999b, Giacalone-and-Jokipii-1999, Qin-etal-2002} so it is assumed that:
\begin{eqnarray}\label{perp1}
K_{\bot r}=aK_{||}\frac{f_{3}(t)}{f_{2}(t)},
\end{eqnarray}
and
\begin{eqnarray}\label{perp2}
K_{\bot \theta}=bK_{||}F(\theta)\frac{f_{3}(t)}{f_{2}(t)},
\end{eqnarray}
where $a$ and $b$ are constants determining the ratio between these coefficients. Here, $F(\theta)$ is a function enhancing $K_{\bot \theta}$ towards the poles by a factor of six \citep{Potgieter-2000, Ferreira-and-Potgieter-2004} and $f_{3}(t)$ is [as given below in Equation (\ref{f3})] a different time-varying function when compared to $f_{2}(t)$ and which incorporates solar cycle related changes into these coefficients. 

Note that, as will be discussed below, $ K_{||}$ and $K_{\bot}$ [\textit{i.e.} $K_{\bot r}$ and $K_{\bot \theta}$] depend differently on magnetic field magnitude and variance and have different functions simulating the time dependence. Therefore, when $K_{\bot}$ is expressed in terms of $K_{||}$ as above, the expression is divided by $f_{2}(t)$ to remove the time dependence of $ K_{||}$ and multiplied by $f_{3}(t)$ to describe the time dependence of $K_{\bot}$.

For the time dependence in the perpendicular diffusion coefficients, the expression for the perpendicular mean free path [$\lambda_{\bot}$] as given by \citet{Shalchi-etal-2004a} is utilised,
\begin{eqnarray}\label{perp}
\lambda_{\bot} \approx \left[{\frac{2\nu-1}{4\nu}} \ F_{2}(\nu) \ a_{k}^{2} \ \frac{\delta{B_{\mathrm{2D}}^2}}{B^2} \sqrt{3}\ l_{\mathrm{2D}} \right]^{\frac{2}{3}}\lambda_{||}^{\frac{1}{3}}.
\end{eqnarray}
Here, $\nu = 5/6$ is the spectral index of magnetic turbulence spectrum, $a_{k} = \frac{1}{\sqrt{3}}$,  $l_{\mathrm{2D}}$ the 2D correlation length, where $l_{\mathrm{2D}} = \frac{l_{\mathrm{slab}}}{10} = 4.55 \times 10^9$m [here $l_{\mathrm{slab}}$ is the slab correlation length] and 
\begin{eqnarray}
F_{2}(\nu)= \sqrt{\pi} \ \frac{\Gamma(\nu)}{\Gamma(\nu-\frac{1}{2})} \frac{2\nu}{2\nu-1}, \nonumber
\end{eqnarray}
where $\Gamma(\nu)=\int_0^\infty x^{\nu-1}e^{-x}\mathrm{d}x$ for $\nu >0$.

Since only the influence of time-varying quantities $B$ and $\delta{B^{2}}$ on $\lambda_{\bot}$ is considered, the expression for $\lambda_{\bot}$ in Equation (\ref{perp}) is approximated as
\begin{eqnarray}\label{perpApprox}
\lambda_{\bot} \propto \left(\frac{\delta{B_{\mathrm{2D}}}}{B}\right)^\frac{4}{3}  \left(\left( \frac{1}{\delta{B_{\mathrm{slab},x}}}\right)^2\left[\frac{1}{4}+\frac{18}{7}\left(\frac{B \ c}{P \ k_{\mathrm{min}}}\right)^{5/3}\right]\right)^\frac{1}{3}.
\end{eqnarray}
From Equation (\ref{perpApprox}) the time dependence for the perpendicular diffusion coefficients, which is described by the function $f_{3}(t)$ in Equations (\ref{perp1}) and (\ref{perp2}), is deduced as
\begin{eqnarray}\label{f3}
f_{3}(t) = C_{5}\left(\frac{\delta{B(t)}}{B(t)}\right)^\frac{4}{3}  \left(\left( \frac{1}{\delta{B(t)}}\right)^2\left[\frac{1}{4}+\frac{18}{7}\left(\frac{B(t) \ c}{P \ k_{\mathrm{min}}}\right)^{5/3}\right] \right)^\frac{1}{3}, 
\end{eqnarray}
with $C_{5}$ a constant in units of $\textrm{(nT)}^{2/3}$.

Concerning drifts, recent theoretical work done by \citet{Minnie-etal-2007b} showed that changes in $\delta B$ over a solar cycle may affect the drift coefficient [$K_{A}$]. As solar activity changes over time so does $\delta B^2$. In this work, a similar dependence for the drift coefficient on solar activity, as in \citet{Minnie-etal-2007b}, is assumed but, instead of $\delta B^2$ , the tilt angle [$\alpha$] is utilised to scale $K_{A}$ for increasing solar activity \citep[see][]{Ferreira-and-Potgieter-2004,Ndiitwani-etal-2005,Manuel-2011a,Manuel-2011b}. This is done to compute realistic charge-sign dependent modulation over a solar cycle as shown by \citet{Ndiitwani-etal-2005}. They showed that $K_{A}$ needs to be scaled even to almost zero for extreme solar maximum periods while for solar minimum periods, $K_{A} \rightarrow 100 \,\%$. This is done by constructing a simple function [$f_1(t)$] that uses $\alpha$ as input parameter, with $\alpha < 75^{\circ}$. The drift coefficient, $K_{A} \propto f_1(t)$ and it is assumed that
\begin{eqnarray}\label{f1}
f_{1}(t)= 0.013 \times \frac{(75.0^{\circ}-\alpha(t))}{\alpha_c},
\end{eqnarray}
with $\alpha_c=1^{\circ}$. See also \citet{Ndiitwani-etal-2005}. Utilising this time dependence, Equation (\ref{eq:KABurger}) is modified as
\begin{equation} \label{Drift2}
K_A = K_{A0}\frac{\beta P}{3B}\frac{10 P^2}{10 P^2+1}f_{1}(t).
\end{equation}

\section{Input Parameters Used in the Model}
\begin{figure}
\begin{center}
\includegraphics*[width=25pc]{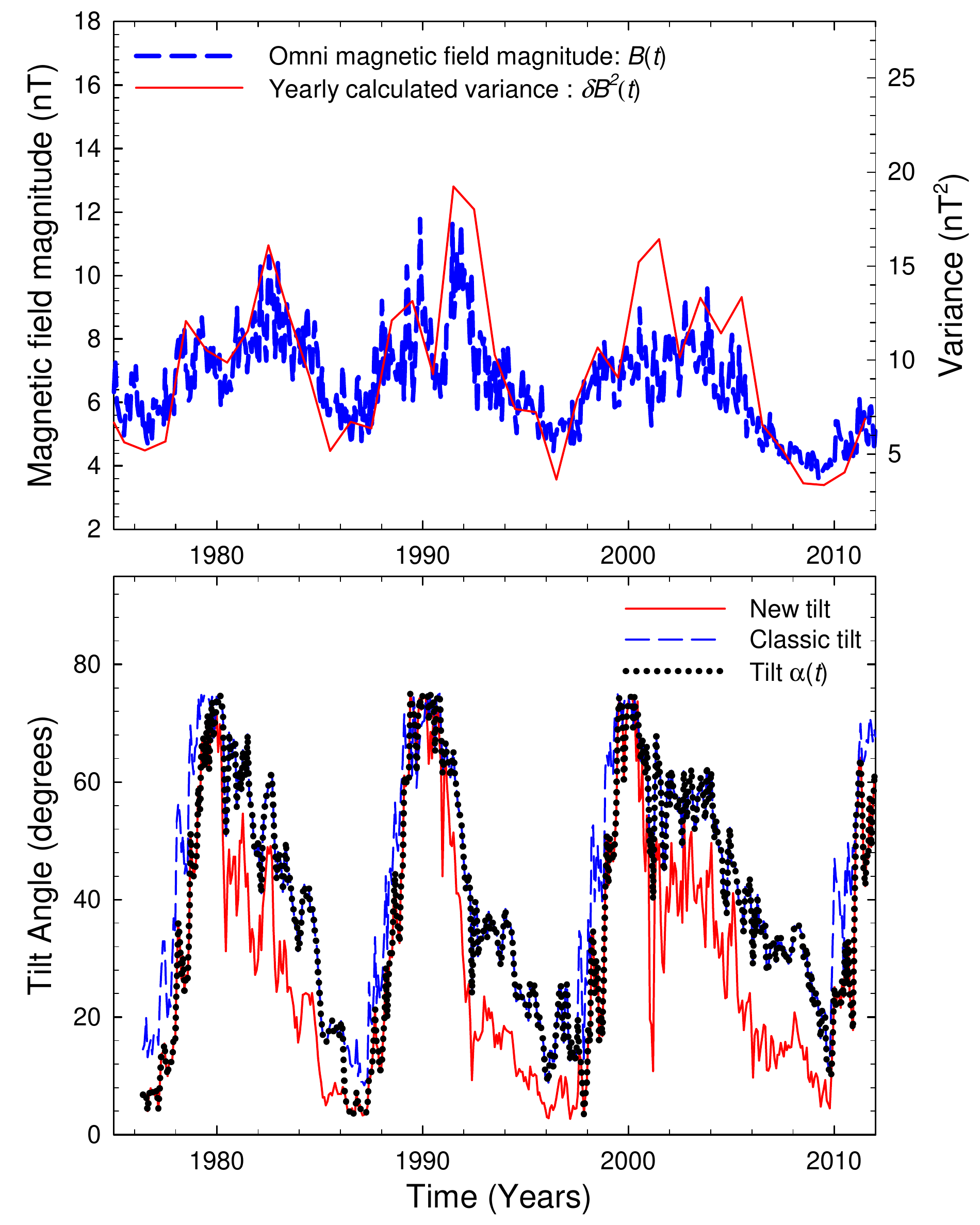}
\end{center}
\caption{Top panel: Shows the observed HMF magnitude at Earth (from NSSDC COHOWeb: \protect\url{nssdc.gfc.nasa.gov/cohoweb}) and  the calculated yearly statistical variance. Bottom panel: Shows the tilt angle as a function of time (see Wilcox Solar Observatory: \protect\url{wso.stanford.edu}) based on two different tilt angle models using different boundary conditions, namely the ``new'' (red solid line) and the ``classic'' (blue dashed line) model \citep[][]{Hoeksema-1992} and the tilt angle used in the model (black dotted line).}
\label{tiltBvar}
\end{figure}
\begin{figure}
\begin{center}
\includegraphics*[width=25pc]{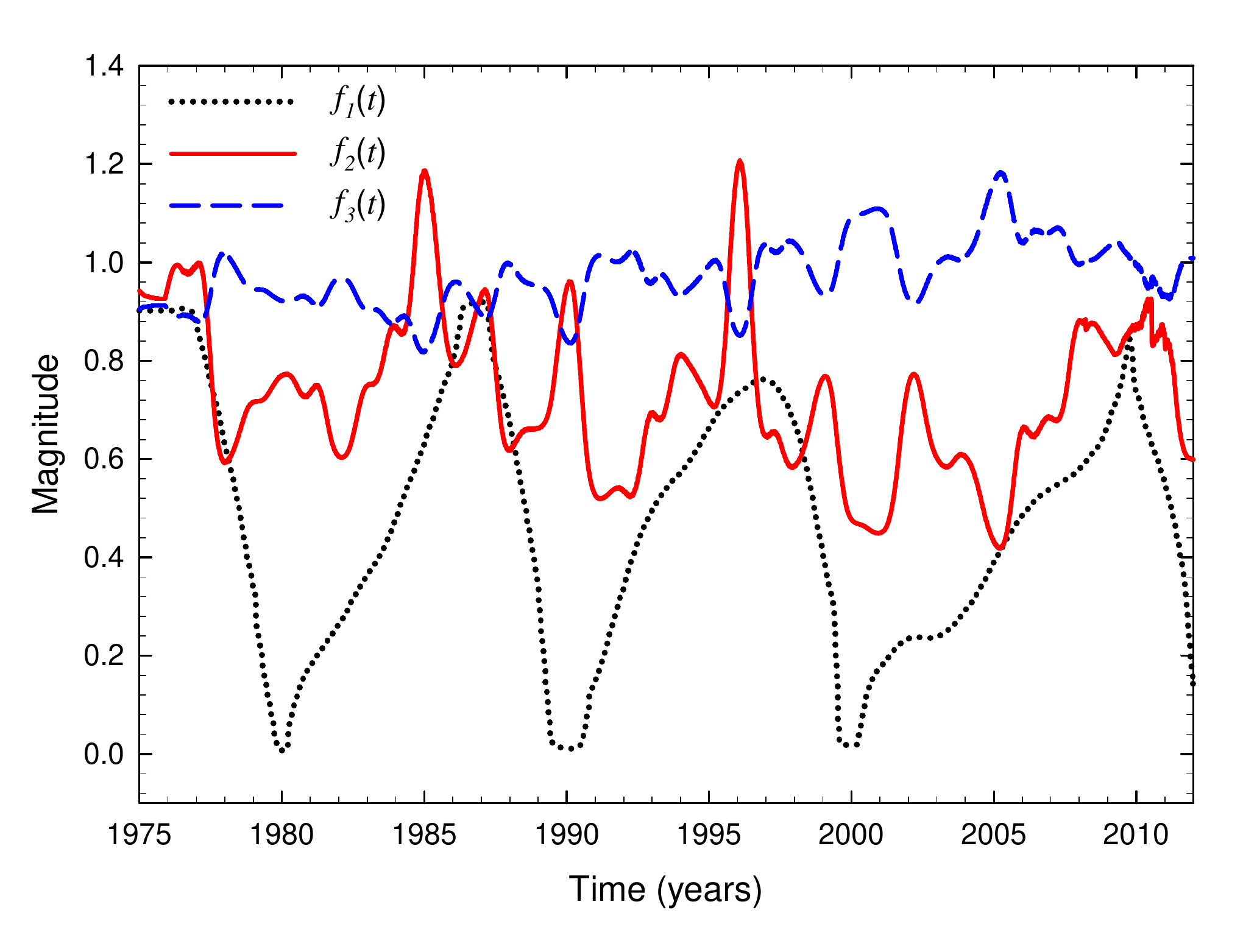}
\end{center}
\caption{The time-dependent functions $f_{1}(t)$ in Equation (\ref{f1}) (dotted black line), $f_{2}(t)$ in Equation (\ref{f2}) (solid red line), and $f_{3}(t)$ in Equation (\ref{f3}) (dashed blue line).}
\label{RTF-TDep}
\end{figure}
To calculate cosmic-ray intensities, the functions $f_{2}(t)$ and $f_{3}(t)$ require $B$ and $\delta{B^{2}}$ as input parameters, which are shown in the top panel of Figure \ref{tiltBvar}. Shown here are the OMNI magnetic field observations until 2012 (from \url{cohoweb.gsfc.nasa.gov}) represented by a blue dashed line.  The total variance [$\delta B^2$] was calculated using the OMNI data. The observed hourly averages of the total field magnitude were binned in one-year intervals, and then the statistical variance in each interval was calculated and is shown as the red solid line in the top panel of Figure \ref{tiltBvar}. These values may change over the termination shock because the magnetic field changes its character after passing through the termination shock.

Shown in the bottom panel of Figure \ref{tiltBvar} is the computed tilt angle \citep{Hoeksema-1992} (see Wilcox Solar Observatory: \url{wso.stanford.edu}) values for two different models using different boundary conditions namely the ``new'' and the ``classic'' model. \citet{Ferreira-and-Potgieter-2003, Ferreira-and-Potgieter-2004} found that the model with the smallest rate of change in $\alpha$ over a period of decreasing or increasing solar activity produce optimal modelling results when compared to cosmic-ray observations.  Taking this into consideration, the ``new'' tilt angle model is used for periods of increasing solar activity while for periods of decreasing solar activity the ``classic'' tilt angle model is used. This is shown in the bottom panel of Figure \ref{tiltBvar}.

The time-dependent function $f_1(t)$ in the drift coefficient, for which $\alpha$ is used as input parameter, is shown in Figure \ref{RTF-TDep} and varies with solar activity, from $\sim$0 during solar maximum to $\sim$0.9 for solar minimum.  Also shown in Figure \ref{RTF-TDep} is the time dependence in the parallel and perpendicular diffusion coefficients, which is produced by the functions $f_{2}(t)$ in Equation (\ref{f2}) and $f_{3}(t)$ in Equation (\ref{f3}) using the input parameters $B$ and $\delta{B^{2}}$ shown in Figure \ref{tiltBvar}. The function $f_{2}(t)$  determines the time-dependent changes in the parallel diffusion coefficient, resulting in a difference between solar minimum and solar maximum by a factor of $\sim$2 while $f_{3}(t)$ is responsible for time-dependent changes in the perpendicular coefficients, only changes by a factor of $\sim$1.2 between solar minimum and maximum.

Because the model is only 2D, all time-dependent effects are transported radially out with the solar wind speed. For solar minimum conditions, this radial speed varies from 400 km\,s$^{-1}$ in the equatorial regions to 800 km\,s$^{-1}$ at the poles while for solar maximum conditions the speed is 400 km\,s$^{-1}$ at all latitudes \citep[see \textit{e.g.}][]{Ferreira-and-Scherer-2006}. After the shock, the radial solar wind speed decreases according to the compression ratio of three \citep{Burlaga-etal-2005,Richardson-etal-2008} and then decreases as 1/$r^2$ further out in the inner heliosheath to the heliopause \citep[\textit{e.g.}][]{Strauss-etal-2010a}.

\citet{Moeketsi-etal-2005} studied the effect of different solar wind speed profiles on the distribution of 7 MeV Jovian and galactic electrons in the inner heliosphere using a 3D steady-state Jovian modulation model. These authors coupled the solar wind speed to the perpendicular diffusion coefficient in the polar direction. They found that the changes in solar wind speed profile from a scenario applicable to solar minimum conditions to one applicable for solar maximum induce changes in the spiral angle  [$\psi$] of the HMF which lead to considerable changes in Jovian electron intensities in the inner heliosphere. Concerning the effect on galactic cosmic rays, a realistic solar wind profile was constructed by \citet{Scherer-and-Ferreira-2005a} and \citet{Ferreira-and-Scherer-2006} using a multi-fluid hydrodynamic model, and it was found that the effect of large changes in the solar wind profile on the cosmic-ray intensities in the inner heliosheath are negligible compared to small changes in the diffusion coefficients.

\section{Modelling Results}
\begin{figure}
\begin{center}
\includegraphics*[width=25pc]{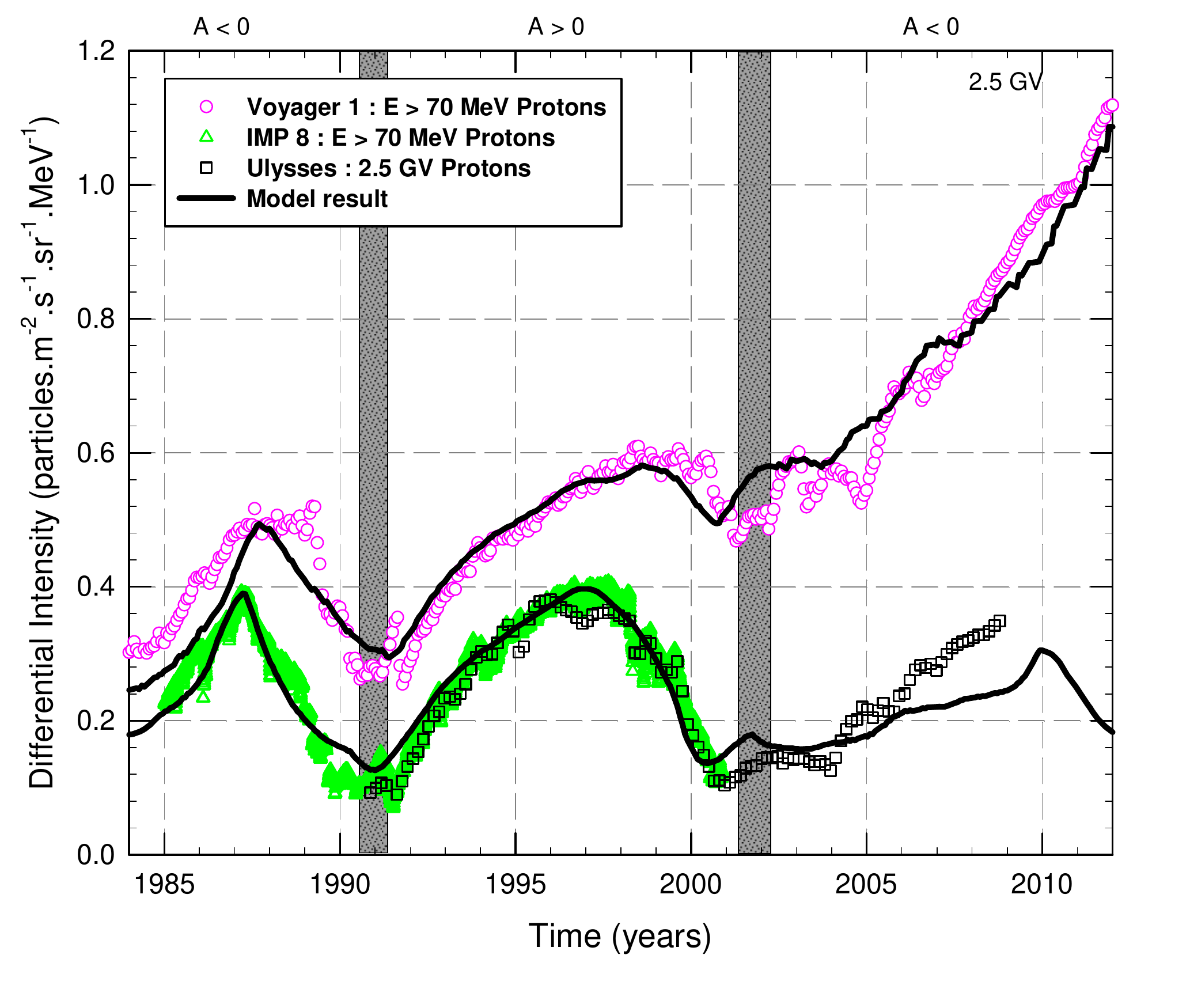}
\end{center}
\caption{Computed 2.5 GV cosmic-ray proton intensities at Earth and along the \textit{Voyager 1} trajectory since 1984 are shown as a function of time. Also shown are the proton observations from \textit{Voyager 1} with $E >$ 70 MeV (from \protect\url{voyager.gsfc.nasa.gov}) as symbols (circles) and for $E > $ 70 MeV measurements at Earth from \textit{IMP 8} (from \protect\url{astro.nmsu.edu}) (triangles) and $\sim$ 2.5 GV proton observations (squares) from \textit{Ulysses} \citep[][]{Heber-etal-2009}. The shaded areas represent the periods when there was not a well defined HMF polarity.}
\label{RTF-data}
\end{figure}

The 2.5 GV proton intensities produced by the 2D time-dependent model, assuming $f_1(t)$, $f_2(t)$, and $f_3(t)$ as in Equations (\ref{f1}), (\ref{f2}) and (\ref{f3}) respectively (shown in Figure \ref{RTF-TDep}) are compared with \textit{Voyager 1}, \textit{International Monitoring Platform (IMP) 8}, and \textit{Ulysses} observations in Figure \ref{RTF-data}. Observed cosmic-ray protons with kinetic energy $E > 70$ MeV as measured by \textit{Voyager 1} and \textit{IMP 8} are shown together with 2.5 GV measurements on board the \textit{Ulysses} spacecraft. As mentioned above, the aim of this work is not to obtain detailed fits to these observations, but rather to establish compatibility between the model and these observations globally. This is done in order to reproduce the modulation amplitude over three consecutive solar cycles at Earth and along the \textit{Voyager 1} trajectory to find whether the recent theory could be used in a model to reproduce, to first-order, long-term cosmic-ray modulation. 

Note that it is expected that the $E > 70$ MeV channel has a substantial contribution from anomalous cosmic-ray (ACRs) protons especially in the inner heliosheath region \citep{Stone-etal-2008} and therefore the observations along the \textit{Voyager 1} trajectory after 2004 should be interpreted as an upper limit. 

Figure \ref{RTF-data} shows the $E >$ 70 MeV proton observations (from \url{voyager.gsfc.nasa.gov}) from the \textit{Voyager 1} spacecraft as a function of time. Also shown are $E > $ 70 MeV measurements at Earth from \textit{IMP 8} (from \url{astro.nmsu.edu}) \citep[\textit{e.g.}][]{Webber-and-Lockwood-1995} and $\sim$ 2.5 GV proton observations from \textit{Ulysses} \citep{Heber-etal-2009}. The \textit{Ulysses} and \textit{IMP 8} observations largely agree on this global scale where they have almost the same modulation amplitude from solar minimum to solar maximum. Note that the \textit{Ulysses} spacecraft did move to higher latitudes and larger distances and therefore cannot be compared in detail with \textit{IMP 8} data without being corrected for latitudinal and radial gradients \citep{Heber-etal-2009}. Since this study is mainly focussed on global modulation over a solar cycle, this data set is used only as an extension of the \textit{IMP 8} data into the recent polarity cycle to give an indication of the modulation amplitude.

In Figure \ref{RTF-data}, computed results are shown corresponding to parameters that are optimised to fit observations in the inner and outer heliosphere. They are as follows: $r_{\mathrm{hp}}=119$ AU, $r_{\mathrm{ts}}=90$ AU [in Equation (\ref{par-gt-rts})], $a=0.014$ [in Equation (\ref{perp1})],  $b=0.01$ [in Equation (\ref{perp2})], $K_{A0}=0.8$ [in Equation (\ref{Drift2})], $C_1=3.0$ [in Equations (\ref{par}) and (\ref{par-gt-rts})], $C_2=0.8$ [in Equations (\ref{par}) and (\ref{par-gt-rts})], $k_{\mathrm{min}} = 10^{-10}$ m$^{-1}$ at 1 AU [in Equations (\ref{f2}) and (\ref{f3})] and the compression ratio = 3.0 [in Equation (\ref{par-gt-rts})]. The proton spectra assumed at higher energies are similar to \cite{Moskalenko-etal-2002} \citep[See also][]{Ptuskin-etal-2006,Webber-and-Higbie-2009} but modified for low energies to fit the \textit{Voyager 1} observations at 119 AU (the distance just before the spacecraft reached the heliosheath depletion region). These spectra are assumed as the heliopause spectra (HPS) at the assumed modulation boundary. 

Figure \ref{RTF-data} illustrates that the model, assuming this set of parameters and time dependence in the coefficients as described above, produces modulation that is compatible with the observations on a global scale. This aspect was also reported on by \cite{Manuel-2011a} and \cite{Manuel-2011b}. Although differences exist for certain periods (especially after $\sim$2004 at Earth) between the model and observations, it can be concluded that the functions $f_1(t)$, $f_2(t)$, and $f_3(t)$ and their particular dependence on $\delta B^2$, $B$, and $\alpha$ result in general compatible modulation over several solar and magnetic cycles and both in the inner and outer heliosphere.

However, differences between the model and observations do exist in Figure \ref{RTF-data} that need to be mentioned. First, towards solar maximum, the computed step decreases are not as pronounced as observed. This indicates that merging of diffusion barriers \citep{Burlaga-etal-1993,Leroux-and-Potgieter-1995} should be included in the model to reproduce these specific periods. This is however beyond the scope of this study. Second, for the period $\sim$1984-$\sim$1987, the model computed lower intensities than observed by \textit{Voyager 1} and for the period $\sim$1988-$\sim$1990, the model is decreasing faster towards solar maximum compared to the observations. This aspect where the model over-estimates the onset of solar maximum was also discussed by \cite{Ferreira-and-Potgieter-2004}.  The most notable differences between the model and observations in Figure \ref{RTF-data} are after $\sim$2004 at Earth. For this period the model results in compatible intensities along the \textit{Voyager 1} trajectory, but at Earth the computed intensities are recovering much more slowly towards solar minimum than observed. This aspect is now further investigated by first illustrating the effect of different parameters on the computed intensities to establish if one of them could lead to improved compatibility.

\subsection{Effect of Different Heliopause Positions}
\begin{figure}
\begin{center}
\includegraphics*[width=25pc]{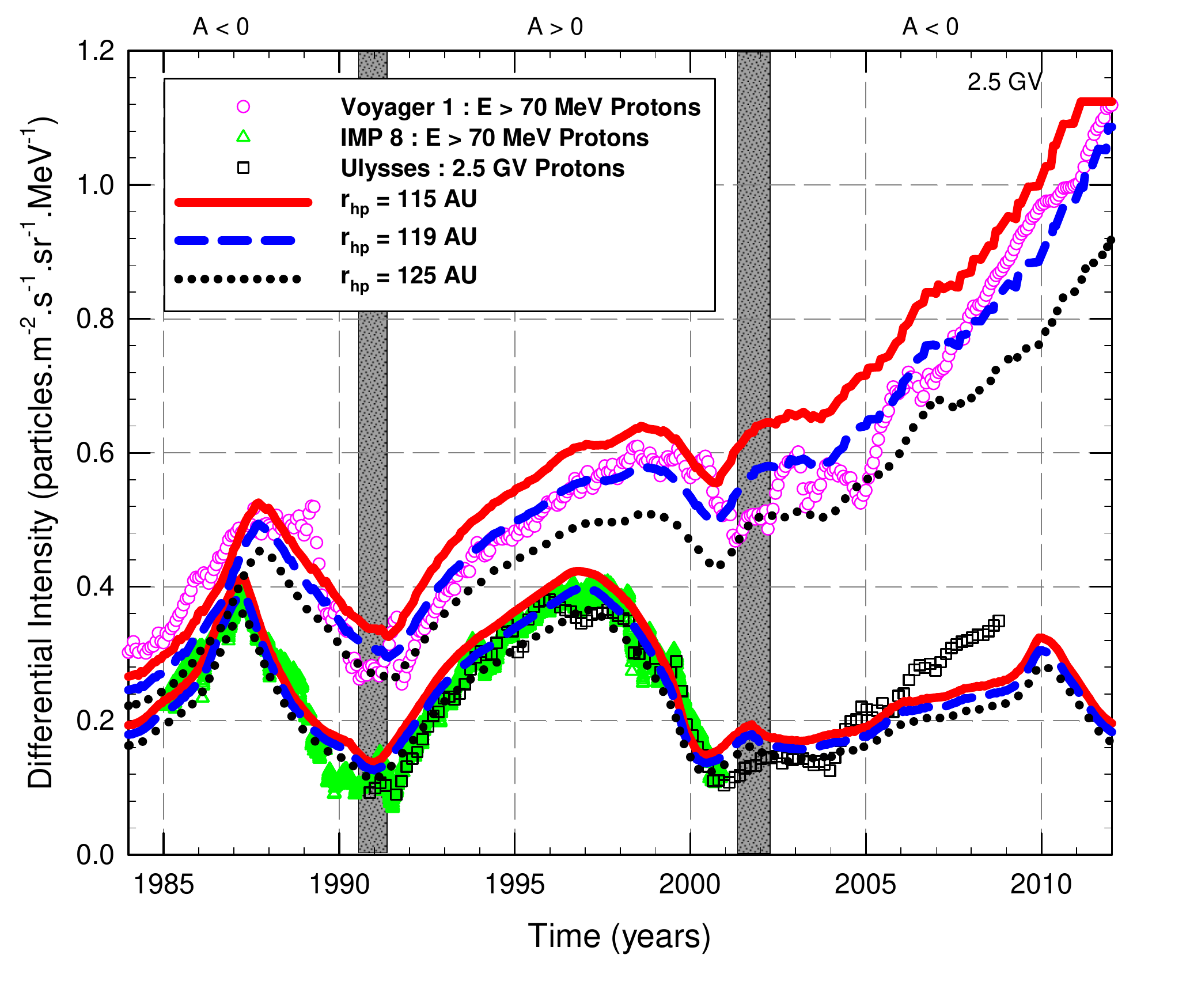}
\end{center}
\caption{Computed 2.5 GV cosmic-ray proton intensities at Earth and along the \textit{Voyager 1} trajectory since 1984 are shown for three different heliopause radii [$r_{\mathrm{hp}}$] as a function of time. Also shown are the $E >$ 70 MeV proton observations from \textit{Voyager 1} (from \protect\url{voyager.gsfc.nasa.gov}) as symbols (circles) and for $E > $ 70 MeV measurements at Earth from \textit{IMP 8} (from \protect\url{astro.nmsu.edu}) (triangles) and $\sim$ 2.5 GV proton observations (squares) from \textit{Ulysses} \citep[][]{Heber-etal-2009}. The shaded areas represent the periods when there was not a well defined HMF polarity.}
\label{RTF-rb}
\end{figure}
The computed cosmic-ray intensities corresponding to different heliopause radii [$r_{\mathrm{hp}}$] are shown in Figure \ref{RTF-rb} in an attempt to find improved compatibility with observations at Earth after $\sim$2004. The discussion of Figure \ref{RTF-rb} starts with the dashed blue line, which is considered as the optimal result compared to the observations and shown in the previous figure. For this model, the modulation boundary is assumed at 119 AU. Note that this choice of the assumed heliopause position in the model is in general accordance to the recent \textit{Voyager 1} observations \citep{Krimigis-etal-2011} where it is predicted as $121_{+16}^{-11}$ AU. Recently \textit{Voyager 1} crossed a region that could be interpreted as the heliopause at 121.7 AU \citep{Stone-etal-2013,Webber-McDonald-2013}. However, for our model we assume the spectra observed by \textit{Voyager 1} just before it reached the heliosheath depletion region \citep{Burlaga-etal-2013} as the HPS due to the limitations in understanding this region. Computations based on assuming a 115 AU and a 125 AU boundary are also shown in Figure \ref{RTF-rb}, in order to illustrate the effect of  much smaller and larger boundary on cosmic-ray modulation. These scenarios are still possible options, until the heliopause is eventually observed by both \textit{Voyager 1} and \textit{2}.

It follows from Figure \ref{RTF-rb} that when the heliopause position is increased from  115 AU to 125 AU and keeping all other parameters in the model unchanged, the calculated cosmic-ray intensities decrease at Earth and especially along the \textit{Voyager 1} trajectory, becoming more pronounced when the spacecraft approaches the boundary. Along the \textit{Voyager 1} trajectory, the $r_{\mathrm{hp}}= 115$ AU scenario gives compatible results for the period $\sim$1986\,--\,1989 but not afterwards when it is much higher than the observations. The 125 AU scenario gives lower intensities than the observations, but is compatible during solar maximum periods. This suggests, from a cosmic-ray perspective, that in order to compute such a larger amplitude the heliopause could be situated further away from the Sun during solar maximum periods.

Comparing the different heliopause scenarios in Figure \ref{RTF-rb}, it follows that a time dependence of $r_{\mathrm{hp}}$ cannot improve compatibility with observations at Earth after $\sim$2004, where the computed intensities are consistently lower than the observations. 
 
\subsection{Effect of Different Termination Shock Positions}
\begin{figure}
\begin{center}
\includegraphics*[width=25pc]{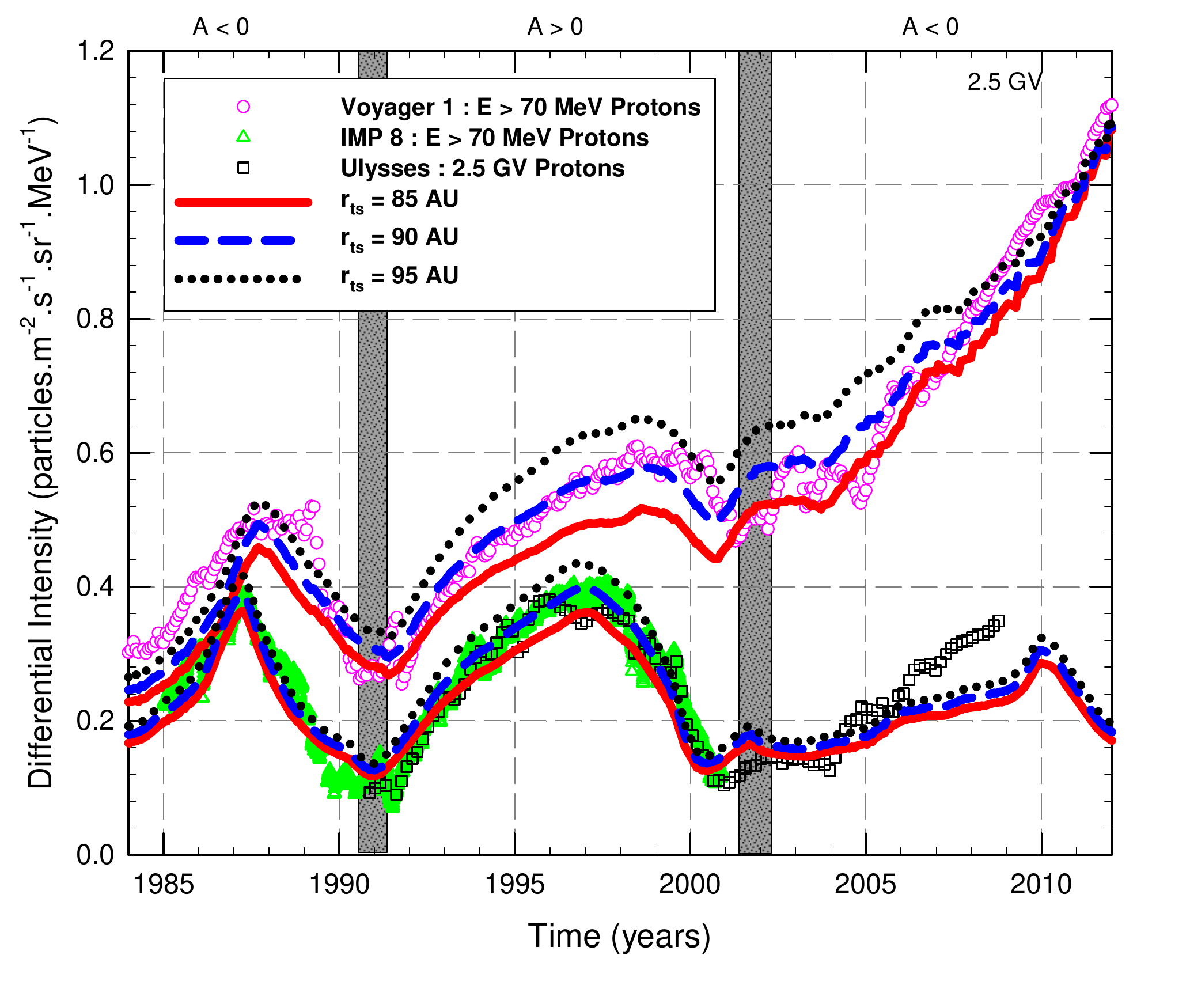}
\end{center}
\caption{Computed 2.5 GV cosmic-ray proton intensities at Earth and along the \textit{Voyager 1} trajectory since 1984 are shown for three different TS positions [$r_{\mathrm{ts}}$] as a function of time. Also shown are the $E >$ 70 MeV proton observations from \textit{Voyager 1} (from \protect\url{voyager.gsfc.nasa.gov}) as symbols (circles) and for $E > $ 70 MeV measurements at Earth from \textit{IMP 8} (from \protect\url{astro.nmsu.edu}) (triangles) and $\sim$ 2.5 GV proton observations (squares) from \textit{Ulysses} \citep[][]{Heber-etal-2009}. The shaded areas represent the periods when there was not a well defined HMF polarity. }
\label{RTF-ts}
\end{figure}
Figure \ref{RTF-ts} shows the computed intensities corresponding to different assumed TS positions [$r_{\mathrm{ts}}$]. Three computed scenarios are shown, namely $r_{\mathrm{ts}}=$ 85 AU, 90 AU and 95 AU respectively, which are between what was observed by \textit{Voyager 1} and \textit{Voyager 2} \citep{Stone-etal-2005,Stone-etal-2008}. For all three scenarios, $r_{\mathrm{hp}}=119$ AU. The scenario with $r_{\mathrm{ts}}=90$ AU (dashed blue line) is considered the reference scenario, which produces the best fit when compared to observations. It is known that the TS position is changing as a function of solar activity \citep{Scherer-and-Fahr-2003b,Webber-and-Intriligator-2011} and is therefore not stationary, as assumed here. The purpose of this section however is to show the sensitivity of the computed intensities to the position of the shock.  

The computed scenarios in Figure \ref{RTF-ts} show that a change in $r_{\mathrm{ts}}$ has a significant effect on cosmic-ray intensities along the \textit{Voyager 1} trajectory when compared to the Earth. Also, it shows that at Earth during solar maximum, the considered scenarios produce almost the same result. However, for solar minimum when $r_{\mathrm{ts}}$ is decreased, the cosmic-ray intensities also decrease. This effect is much more prominent along the \textit{Voyager 1} trajectory, where it is shown that when the thickness of the inner heliosheath is decreased (\textit{e.g.} increasing $r_{\mathrm{ts}}$ but keeping  $r_{\mathrm{hp}}$ the same), cosmic-ray intensities in the heliosphere as a whole increase. Although there is no acceleration of cosmic rays considered in this model, the effect of the inner heliosheath is simulated by changing (decreasing) the transport parameters across the TS. This means that the inner heliosheath acts as a modulation barrier \citep{Potgieter-and-Leroux-1989,Ferreira-etal-2004,Langner-etal-2004,Ngobeni-Potgieter-2010,Ngobeni-2011,Potgieter-Nndanganeni-2013} and increasing the thickness results in fewer cosmic rays entering the rest of heliosphere. Figure \ref{RTF-ts} shows that a time dependence in $r_{\mathrm{ts}}$ is also not the answer to compute compatible results for the intensities after $\sim$2004, but may give improved compatibility during other periods. In addition, although not shown, changing the compression ratio [$s_k$] causing larger or smaller decreases in the diffusion coefficient in the model also fails to reproduce compatible intensities at Earth after $\sim$2004. 

\begin{figure}
\begin{center}
\includegraphics*[width=25pc]{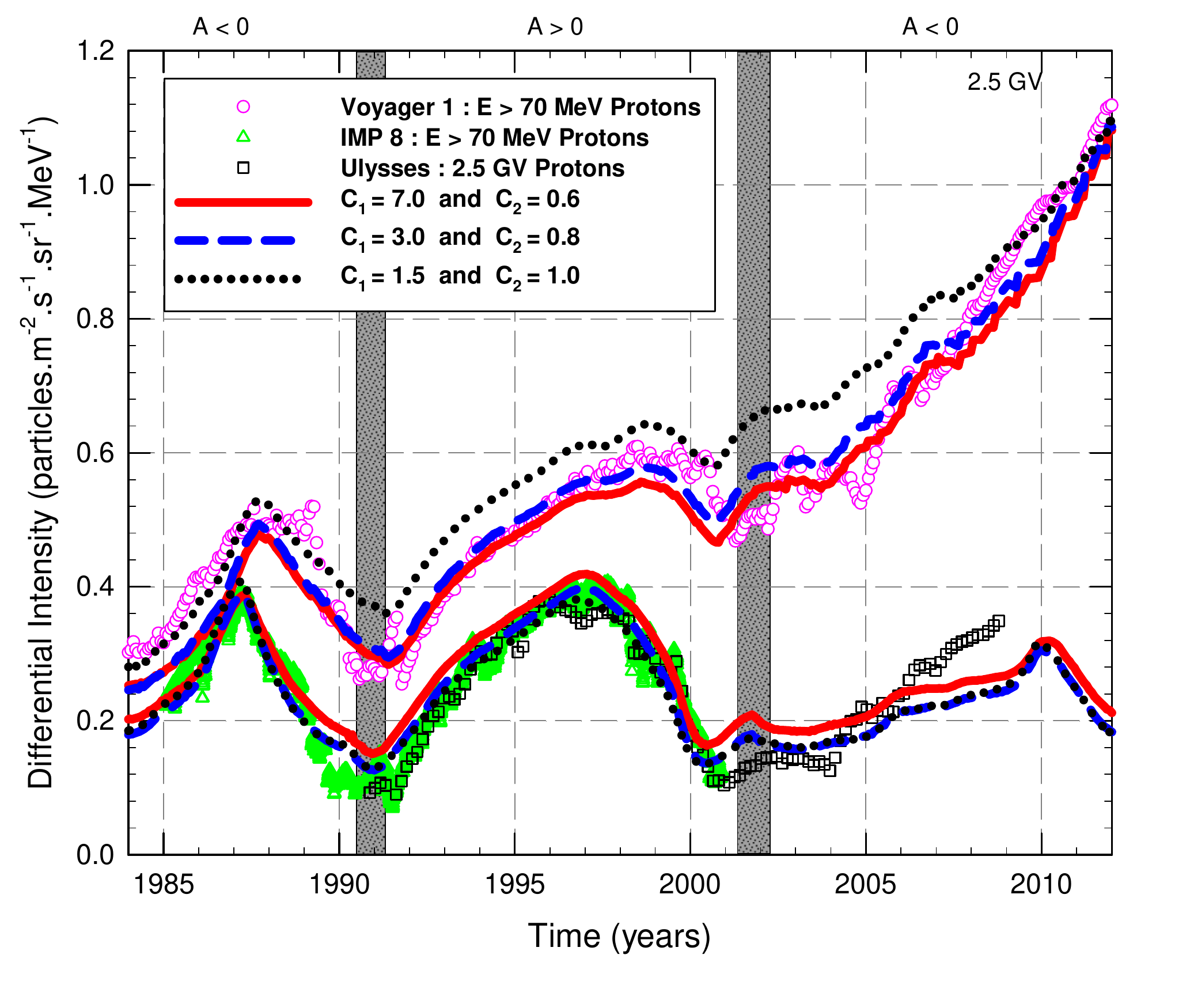}
\end{center}
\caption{Computed 2.5 GV cosmic-ray proton intensities at Earth and along the \textit{Voyager 1} trajectory since 1984 are shown for three different $C_1$ and $C_2$ values, as given in Equations (\ref{par}) and (\ref{par-gt-rts}), as a function of time. Also shown are the $E >$ 70 MeV proton observations from \textit{Voyager 1} (from \protect\url{voyager.gsfc.nasa.gov}) as symbols (circles) and for $E > $ 70 MeV measurements at Earth from \textit{IMP 8} (from \protect\url{astro.nmsu.edu}) (triangles) and $\sim$ 2.5 GV proton observations (squares) from \textit{Ulysses} \citep[][]{Heber-etal-2009}. The shaded areas represent the periods when there was not a well defined HMF polarity. }
\label{RTF-dpa}
\end{figure}

\subsection{Effect of Different $C_1$ and  $C_2$ Values}
The effect of different $C_1$ and $C_2$ values, as given in Equations (\ref{par}) and (\ref{par-gt-rts}), on computed intensities is shown in Figure \ref{RTF-dpa}.  The $C_1$ value determines the magnitude of the transport coefficients at Earth and $C_2$ the radial dependence, where $K_{||} \propto C_1 r^{C_2}$.  From the figure, it follows that when the radial dependence is increased by changing $C_2 =0.6$ to 1.0, the $C_1$ value has to be decreased from 7.0 to 1.5 to reduce the computed intensities at Earth to become compatible to observations. This however increases the intensity along the \textit{Voyager 1} trajectory. Also, it is evident that these different coefficient scenarios fit the observations better for different solar cycles, \textit{e.g.} during 1985\,--\,1990 ($A<0$ cycle), $C_1=1.5$ and $C_2=1.0$ give a better fit compared to observations than the other two scenarios, but during 1991\,--\,2001 ($A>0$ cycle), $C_1=3.0$ and $C_2=0.8$ gives a better fit to observations. This suggests that different diffusion coefficients 
can be assumed for different polarity cycles \citep[\textit{e.g.}][]{Reinecke-etal-1996,Burger-etal-2000,Potgieter-2000,Ferreira-and-Potgieter-2004} to improve compatibility with the observations. However, $C_1=3.0$ and $C_2=0.8$  are considered as the overall best compatible scenario. Again, after $\sim$2004, all of the different scenarios fail to reproduce the observations at Earth. 

\subsection{Effect of Different $a$ and $b$ Values}
\begin{figure}
\begin{center}
\includegraphics*[width=25pc]{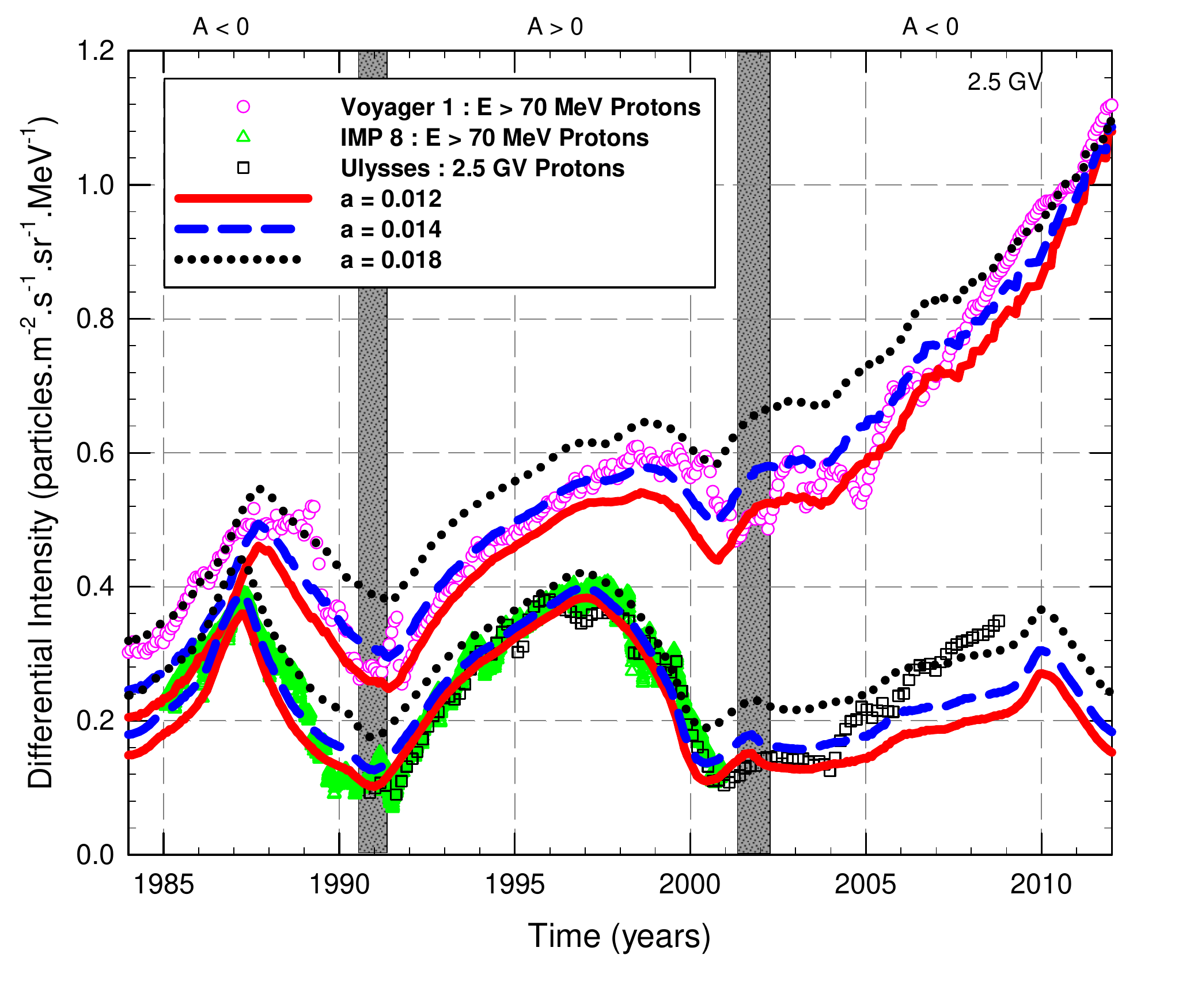}
\end{center}
\caption{Computed 2.5 GV cosmic-ray proton intensities at Earth and along the \textit{Voyager 1} trajectory since 1984 are shown for three different $a$ values, where $a$ is the ratio of perpendicular diffusion coefficient in radial direction to the parallel diffusion coefficient as given by Equation (\ref{perp1}), as a function of time. Also shown are the $E >$ 70 MeV proton observations from \textit{Voyager 1} (from \protect\url{voyager.gsfc.nasa.gov}) as symbols (circles) and for $E > $ 70 MeV measurements at Earth from \textit{IMP 8} (from \protect\url{astro.nmsu.edu}) (triangles) and $\sim$ 2.5 GV proton observations (squares) from \textit{Ulysses} \citep[][]{Heber-etal-2009}. The shaded areas represent the periods when there was not a well defined HMF polarity. }
\label{RTF-a-value}
\end{figure}
The effects on computed intensities of the ratio of perpendicular diffusion coefficient in the radial direction to the parallel diffusion coefficient [$a$] as given in Equation (\ref{perp1}), is shown in Figure \ref{RTF-a-value}. Three scenarios corresponding to $a=$ 0.012, 0.014, and 0.018 in the model are shown with the value 0.014 considered as the optimal value when compared to observations. When the $a$-value is increased, the cosmic-ray intensities generally increase because of the larger perpendicular diffusion coefficient. This is evident at Earth and along the \textit{Voyager 1} trajectory. When compared to observations, for the period 1985\,--\,1989 during an $A<0$ polarity cycle, $a=0.018$  produces an optimal result when compared to observations. However, when compared to observations during solar maximum periods, a smaller $a$ value (0.012) produces the more compatible result. For the reference scenario ($a=0.014$), the model produces a global compatibility at Earth until $\sim$2004, but not thereafter. By increasing the $a$-value to 0.018, the intensities tend to increase to observed solar minimum values at Earth but may still be too high along the \textit{Voyager 1} trajectory.  These results however suggest that a time-dependent $a$-value may produce better compatibility after $\sim$2004 at Earth. This suggests a different time dependence rather than what is assumed so far, and is investigated next. It is also found that the model also fails to reproduce the observations at Earth for different $b$-values after $\sim$2004.

\begin{figure}
\begin{center}
\includegraphics*[width=25pc]{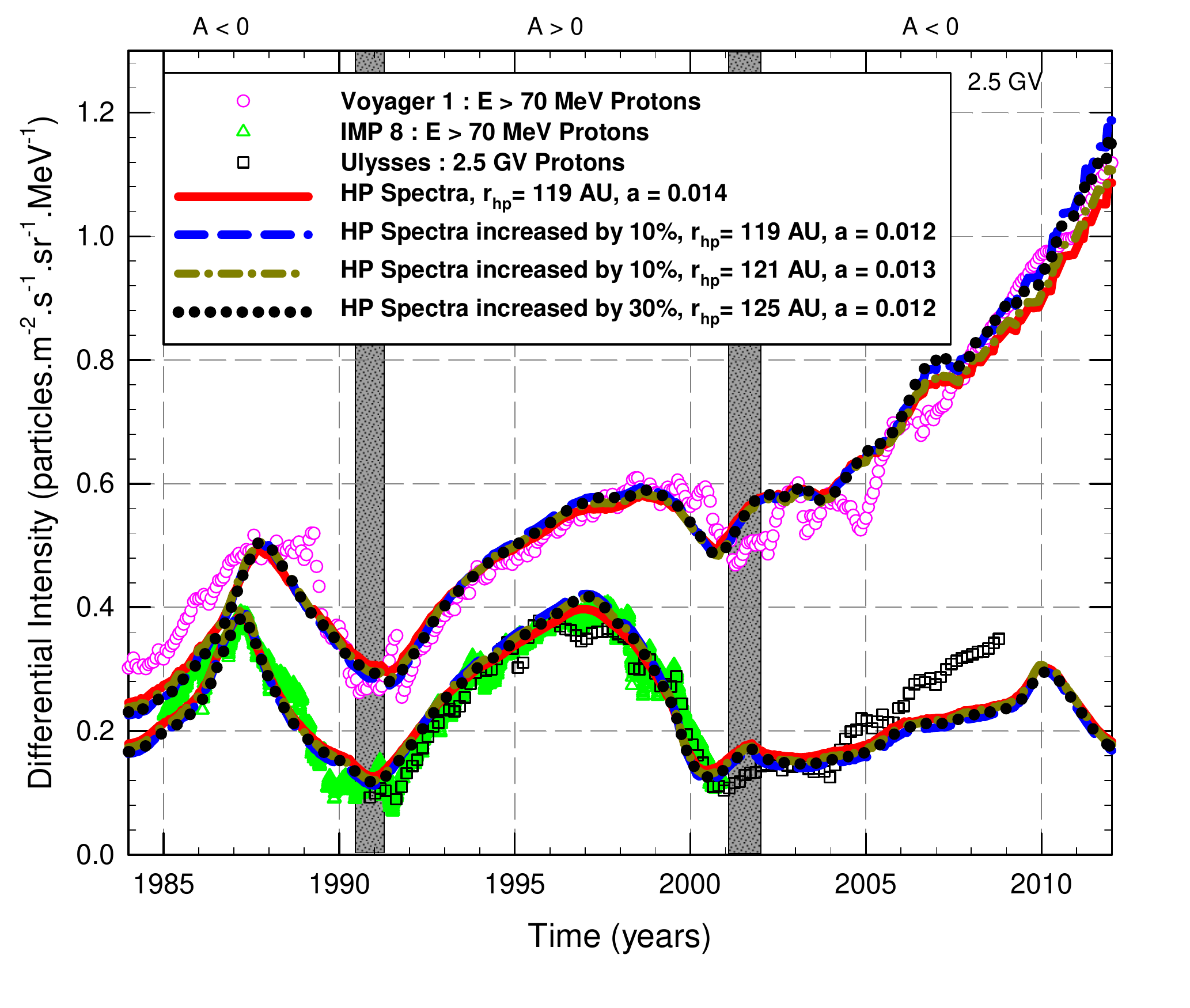}
\end{center}
\caption{Computed 2.5 GV cosmic-ray proton intensities at Earth and along the \textit{Voyager 1} trajectory since 1984 are shown for the assumed HPS values with a 10\,\% and 30\,\% increase and for different $a$-values, as given by Equation (\ref{perp1}), and $r_{\mathrm{hp}}$. Also shown are the $E >$ 70 MeV proton observations from \textit{Voyager 1} (from \protect\url{voyager.gsfc.nasa.gov}) as symbols (circles) and for $E > $ 70 MeV measurements at Earth from \textit{IMP 8} (from \protect\url{astro.nmsu.edu}) (triangles) and $\sim$ 2.5 GV proton observations (squares) from \textit{Ulysses} \citep[][]{Heber-etal-2009}. The shaded areas represent the periods when there was not a well defined HMF polarity.  }
\label{RTF-LIS}
\end{figure}

\subsection{Effect of Different Heliopause Spectra}

For this study, a HPS for protons is assumed based on the 133\,--\,242 MeV and $E>70$ MeV ($\sim$2.5 GV) measurements by \textit{Voyager 1} at $\sim$119 AU. These intensity values are specified at the modulation boundary as the input spectrum to the model. Note that possible modulation in the heliosheath depletion region \cite[][]{Burlaga-etal-2013} and possible additional modulation beyond the heliopause as reported on by \cite{Scherer-et-al-2011} and \cite{Strauss-etal-2013} are not considered. In order to illustrate the effects of a possible higher HPS the assumed values, as in all previous figures, were increased by 10\,\% and then by 30\,\% respectively. 

Figure \ref{RTF-LIS} illustrates four computed intensity scenarios. The solid red line represents the reference scenario with the assumed HPS at 119 AU and $a=0.014$. The dashed blue line represents the scenario where the assumed HPS is increased by 10\,\%, with $r_{\mathrm{hp}}=119$ AU and $a=0.012$ to find compatibility. The computed results in this case follow the reference scenario until $\sim$2010. After $\sim$2010, this scenario produces higher intensities along the \textit{Voyager 1} trajectory. The third scenario is represented by the dash--dotted yellow line, where the assumed HPS values are increased by 10\,\% but with $r_{\mathrm{hp}}=121$ AU and $a=0.013$. The computed intensities for this scenario also follow the reference scenario. A fourth scenario represented by a dotted black line with a 30\,\% increased HPS, $r_{\mathrm{hp}}=125$ AU, and $a=0.012$, also follows the reference scenario. From this it follows that an increase in the HPS values compared to the reference scenario could also result in realistic cosmic-ray modulation if the heliopause position is increased while changing the diffusion coefficients. Therefore, differences from what is assumed in this work as an HPS will not lead to different conclusions because it can easily be off-set by different modulation parameters. Not knowing the exact modulation parameters makes estimates of the true HPS difficult.

\subsection{Effect of Different $K_{A0}$ Values}
\begin{figure}
\begin{center}
\includegraphics*[width=25pc]{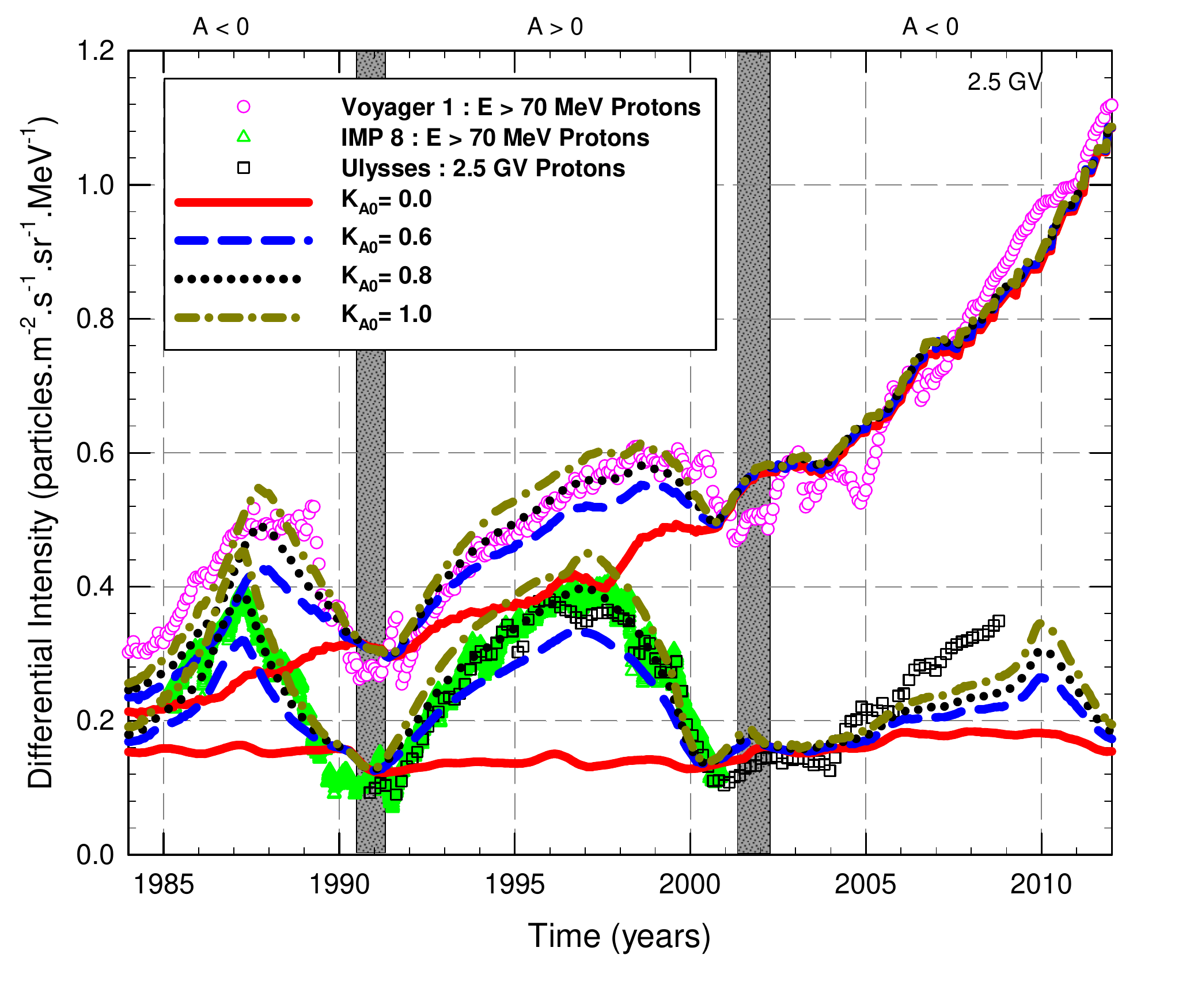}
\end{center}
\caption{Computed 2.5 GV cosmic-ray proton intensities at Earth and along the \textit{Voyager 1} trajectory since 1984 are shown for different $K_{A0}$ values, representing different drift coefficient values as given in Equation (\ref{Drift2}).  Also shown are the $E >$ 70 MeV proton observations from \textit{Voyager 1} (from \protect\url{voyager.gsfc.nasa.gov}) as symbols (circles) and for $E > $ 70 MeV measurements at Earth from \textit{IMP 8} (from \protect\url{astro.nmsu.edu}) (triangles) and $\sim$ 2.5 GV proton observations (squares) from \textit{Ulysses} \citep[][]{Heber-etal-2009}. The shaded areas represent the periods when there was not a well defined HMF polarity. }
\label{RTF-cdmf}
\end{figure}
Figure \ref{RTF-cdmf} shows computed intensities corresponding to different $K_{A0}$ values as given in Equation (\ref{Drift2}), which  scales the drift coefficient. In this figure, four scenarios are shown where $K_{A0}=1.0$ represents a full-drift scenario and $K_{A0}=0.0$ represents a no-drift scenario. Note that all coefficients still change over a solar cycle via Equations (\ref{f1}), (\ref{f2}) and (\ref{f3}) respectively. For extreme solar maximum periods, $K_{A}$ is almost zero via Equation (\ref{f1}), resulting in nearly the same solutions for all $K_{A0}$ scenarios for this level of solar activity. The $K_{A0}=1.0$ scenario gives maximum drift effects for solar minimum, which in turn leads to a maximum cosmic-ray intensities during solar minimum periods. When $K_{A0}$ is decreased from 1.0 to 0.8, then to 0.6, and finally to 0.0, the intensities are also decreasing during solar minimum. For $K_{A0}=0.0$, the cyclic behaviour essentially disappears with only reasonable intensities at solar maximum periods. Note that $K_{A0}=0.8$ is considered as an optimal value when comparing computations to observations along the \textit{Voyager 1} trajectory and at Earth until $\sim$2004. Again after 2004, the model disagrees with the observations at Earth even for a maximum-drift scenario. 

It follows that the difference between solar minimum and solar maximum is largely dependent on the magnitude of the drift coefficient given that the diffusion coefficients have the time dependence as described here. The $K_{A0}=0.8$ assumption gives mostly compatible intensities. When $K_{A0}$ is reduced, the computed amplitude between solar minimum and maximum decreases. This suggests that the computed time dependence may be dominated by solar-cycle related changes in the drift coefficient \citep{Ndiitwani-etal-2005}, as shown in Figure \ref{RTF-cdmf}, where the solid red line (zero drift) shows almost no variation over a solar cycle. However, as will be shown below, the failure of the model to reproduce compatible cosmic-ray intensities at Earth when compared to observations after $\sim$2004 indicates that the assumption of the time dependence in the transport parameters, as given by Equations (\ref{f1}), (\ref{f2}) and (\ref{f3}), is not optimal for this particular solar cycle. This aspect is discussed next. 

\section{Modifying Time Dependence}
After a thorough parameter study as presented above, it is found that when $\delta B^2$, $B$, and $\alpha$ (as shown in Figure \ref{tiltBvar}) are used as time-varying input parameters [for Equations (\ref{f2}), (\ref{f3}), and (\ref{f1})], the model  successfully reproduces cosmic-ray observations on a global scale  along the \textit{Voyager 1} trajectory and at Earth until $\sim$2004, but fails to reproduce reasonable cosmic-ray modulation at Earth after $\sim$2004. It is also shown above that modulation over a solar cycle, computed using these parameters, is dominated by time-dependent changes in the drift coefficient. As will be shown below, this may not be the case for all modulation cycles. However, in a first attempt to compute compatibility with observations at Earth after $\sim$2004, the time dependence in the drift coefficient is modified by constructing a new time-dependent function to replace $f_1(t)$ in Equation (\ref{Drift2}).
\begin{figure}
\begin{center}
\includegraphics*[width=25pc]{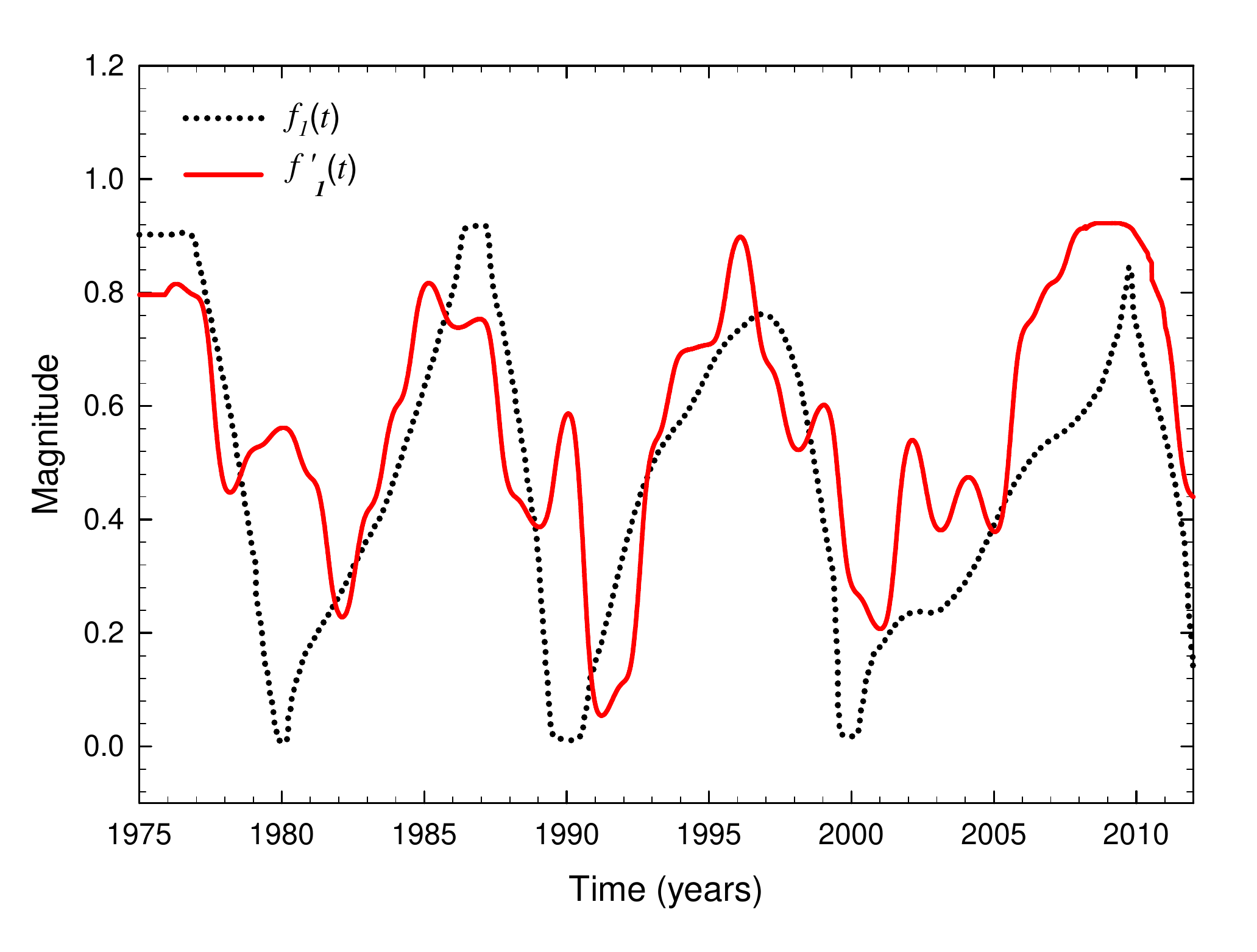}
\end{center}
\caption{Time-dependent drift function [$f_{1}(t)$] compared to the modified time-dependent drift function [$f'_{1}(t)$].}
\label{f1TimeDep}
\end{figure}
\begin{figure}
\begin{center}
\includegraphics*[width=25pc]{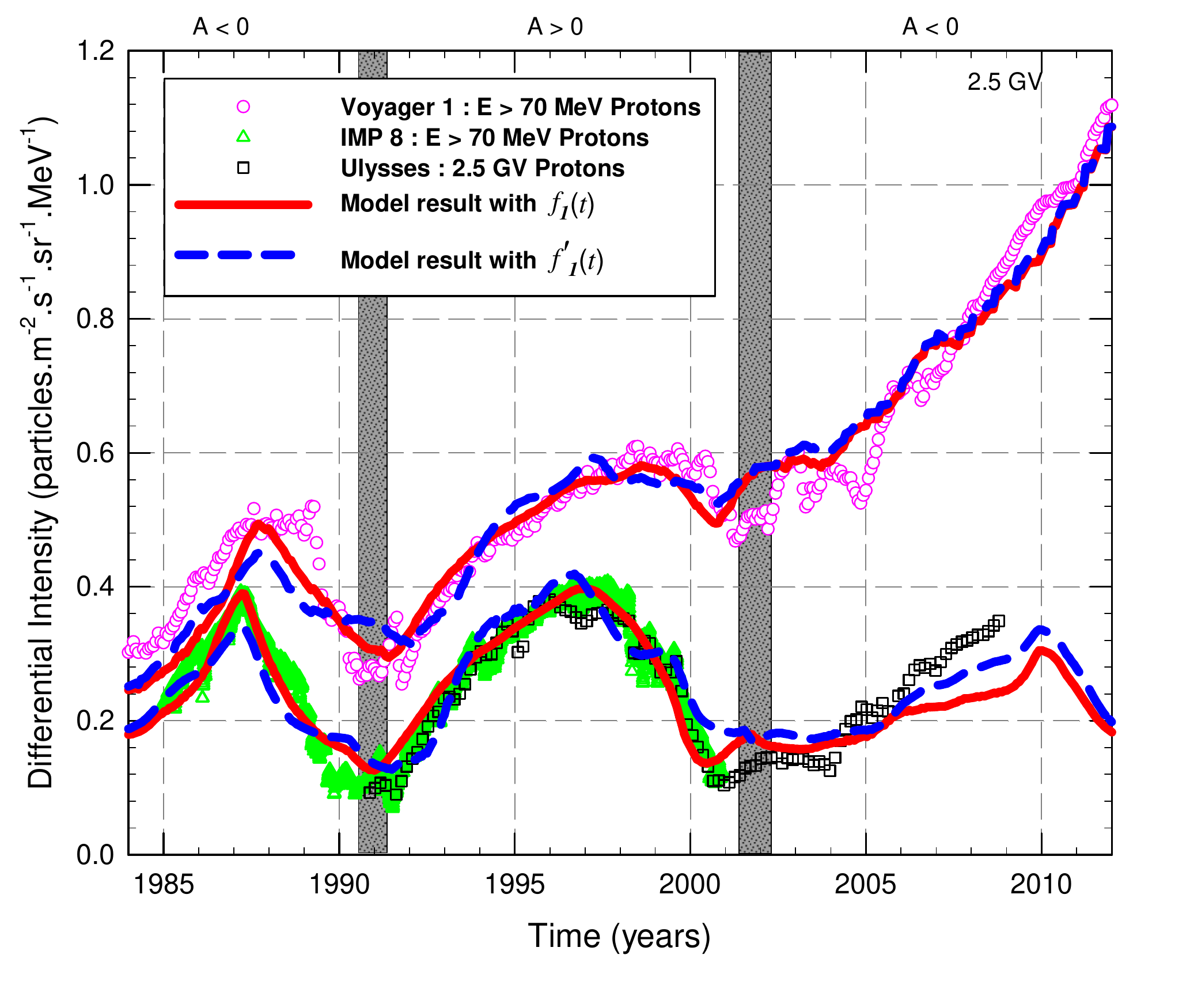}
\end{center}
\caption{Computed 2.5 GV cosmic-ray proton intensities at Earth and along the \textit{Voyager 1} trajectory since 1984 are shown for $f_{1}(t)$ and the changed $f'_{1}(t)$. Also shown are the $E >$ 70 MeV proton observations from \textit{Voyager 1} (from \protect\url{voyager.gsfc.nasa.gov}) as symbols (circles) and for $E > $ 70 MeV measurements at Earth from \textit{IMP 8} (from \protect\url{astro.nmsu.edu}) (triangles) and $\sim$ 2.5 GV proton observations (squares) from \textit{Ulysses} \citep[][]{Heber-etal-2009}. The shaded areas represent the periods when there was not a well defined HMF polarity.}
\label{RTF-f1}
\end{figure}
\subsection{Modifying the Time Dependence in the Drift Coefficient}
To construct a different time-dependent function [$f'_1(t)$] the comparison between the model and observations at Earth after $\sim$2004 is used. From this, it follows that the observed intensities are increasing faster compared to the computed intensities as a function of decreasing solar activity. A function is therefore needed that allows the recovery of drifts earlier compared to the current function as solar activity is decreasing. The function $f_1(t)$ in Equation (\ref{f1}) is now modified. Whereas $f_1(t)$ uses the tilt angle as the only input parameter, the modified function uses the variance [$\delta{B}^2$] \citep{Minnie-etal-2007b}. Different expressions were examined with an optimal expression for $f'_1(t)$ given as
\begin{eqnarray}\label{f1b}
f'_{1}(t)= 1.106-\frac{0.055 \delta{B}^2(t)}{\delta{B_o}^2},
\end{eqnarray}
with $\delta{B_o}^2=1$ nT$^2$.

A comparison between $f_1(t)$ and $f'_1(t)$ is shown in Figure \ref{f1TimeDep}, illustrating that there is a phase difference between them caused by their dependence on different input parameters. However, more important is that for the period from $\sim$2004 onwards, the $f'_1(t)$ is increasing much faster than $f_1(t)$ as a function of decreasing activity. As a matter of fact, this function causes drifts to recover almost immediately to full drifts after $\sim$2004 and should therefore give more realistic cosmic-ray intensities after  $\sim$2004, if the time dependence in this coefficient indeed dominates the recovery of intensities to solar minimum values.

Figure \ref{RTF-f1} depicts the computed cosmic-ray intensities using $f_1(t)$ and $f'_1(t)$ in the model. Shown is that overall $f_1(t)$ gives better compatibility when compared to the modified function $f'_1(t)$. However, for the period from $\sim$2004 onwards at Earth, the modified function produces higher intensities but still much lower than the observations so that the desired recovery of cosmic-ray intensities toward solar minimum in 2009 is not achieved. It is thus concluded that a modification in the time dependence of the diffusion coefficients (as given by $f_2(t)$ and $f_3(t)$) also seems needed, which is discussed next.

\subsection{Modifying the Time Dependence in the Diffusion Coefficients}
\begin{figure}
\begin{center}
\includegraphics*[width=25pc]{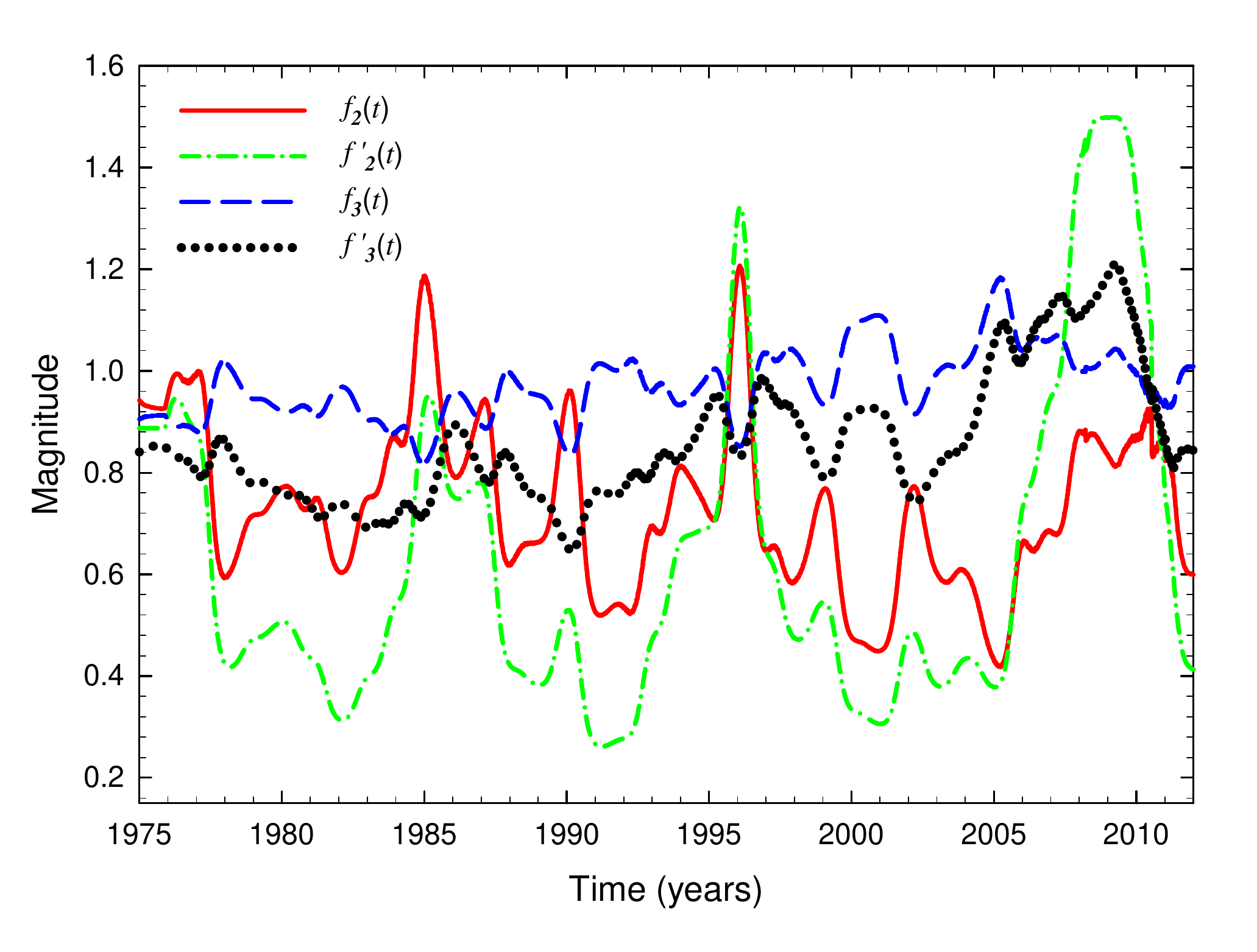}
\end{center}
\caption{The parallel and perpendicular time-dependent functions $f_{2}(t)$ and $f_{3}(t)$ are as given in Equations (\ref{f2}) and (\ref{f3}) compared to the modified functions $f'_{2}(t)$ and $f'_{3}(t)$ as given in Equations (\ref{f2new}) and (\ref{f3new}).}
\label{RTHE-TimeDep}
\end{figure}
\begin{figure}
\begin{center}
\includegraphics*[width=25pc]{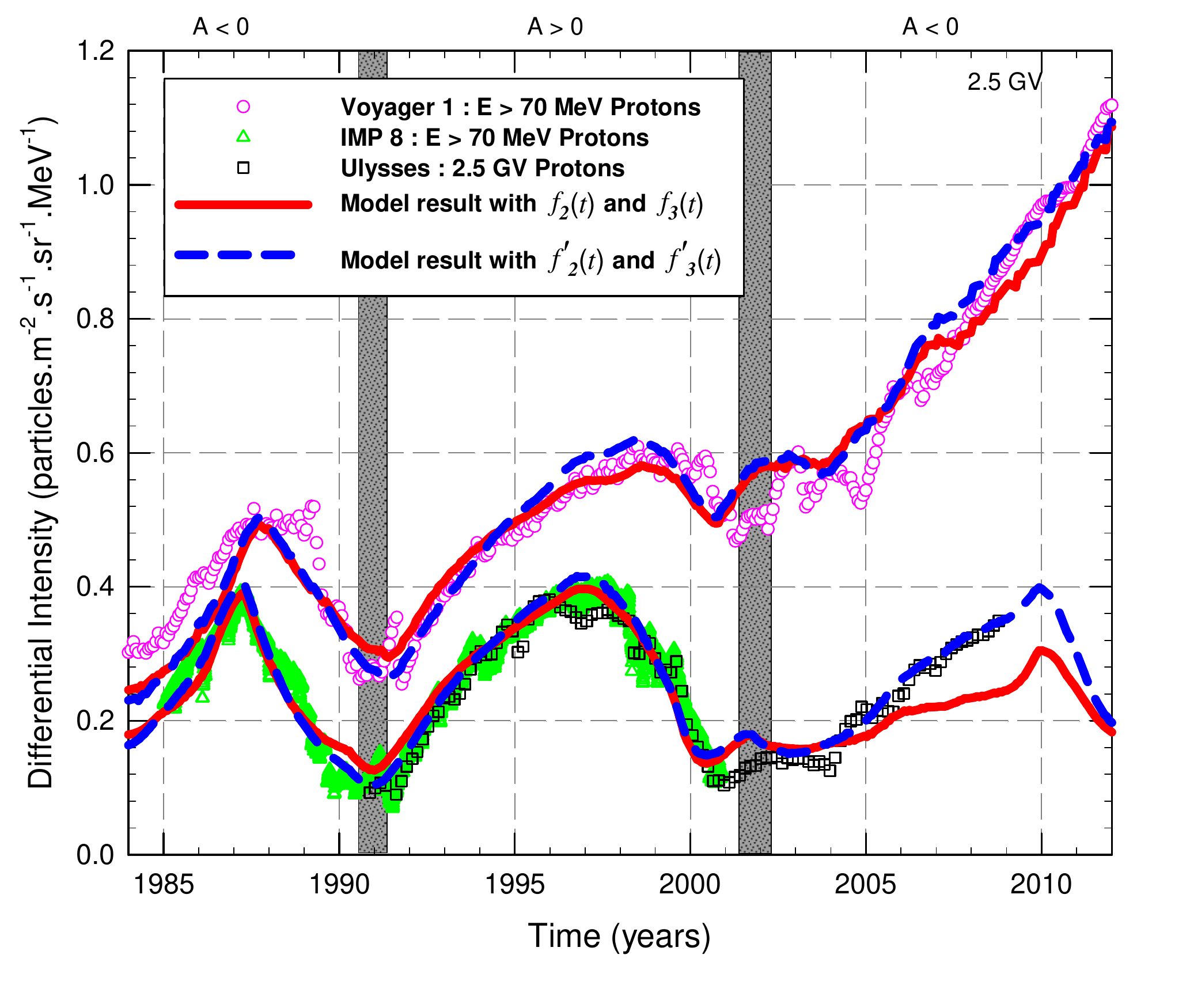}
\end{center}
\caption{Computed 2.5 GV cosmic-ray proton intensities at Earth and along the \textit{Voyager 1} trajectory since 1984 are shown for time-dependent functions $f_{2}(t)$ and $f_{3}(t)$ and the modified time-dependent functions $f'_{2}(t)$ and $f'_{3}(t)$ as given in Equations (\ref{f2}), (\ref{f3}), (\ref{f2new}) and (\ref{f3new}). Also shown are the $E >$ 70 MeV proton observations from \textit{Voyager 1} (from \protect\url{voyager.gsfc.nasa.gov}) as symbols (circles) and for $E > $ 70 MeV measurements at Earth from \textit{IMP 8} (from \protect\url{astro.nmsu.edu}) (triangles) and $\sim$ 2.5 GV proton observations (squares) from \textit{Ulysses} \citep[][]{Heber-etal-2009}. The shaded areas represent the periods when there was not a well defined HMF polarity.}
\label{RTF-RTHE}
\end{figure} 
\begin{figure}
\begin{center}
\includegraphics*[width=25pc]{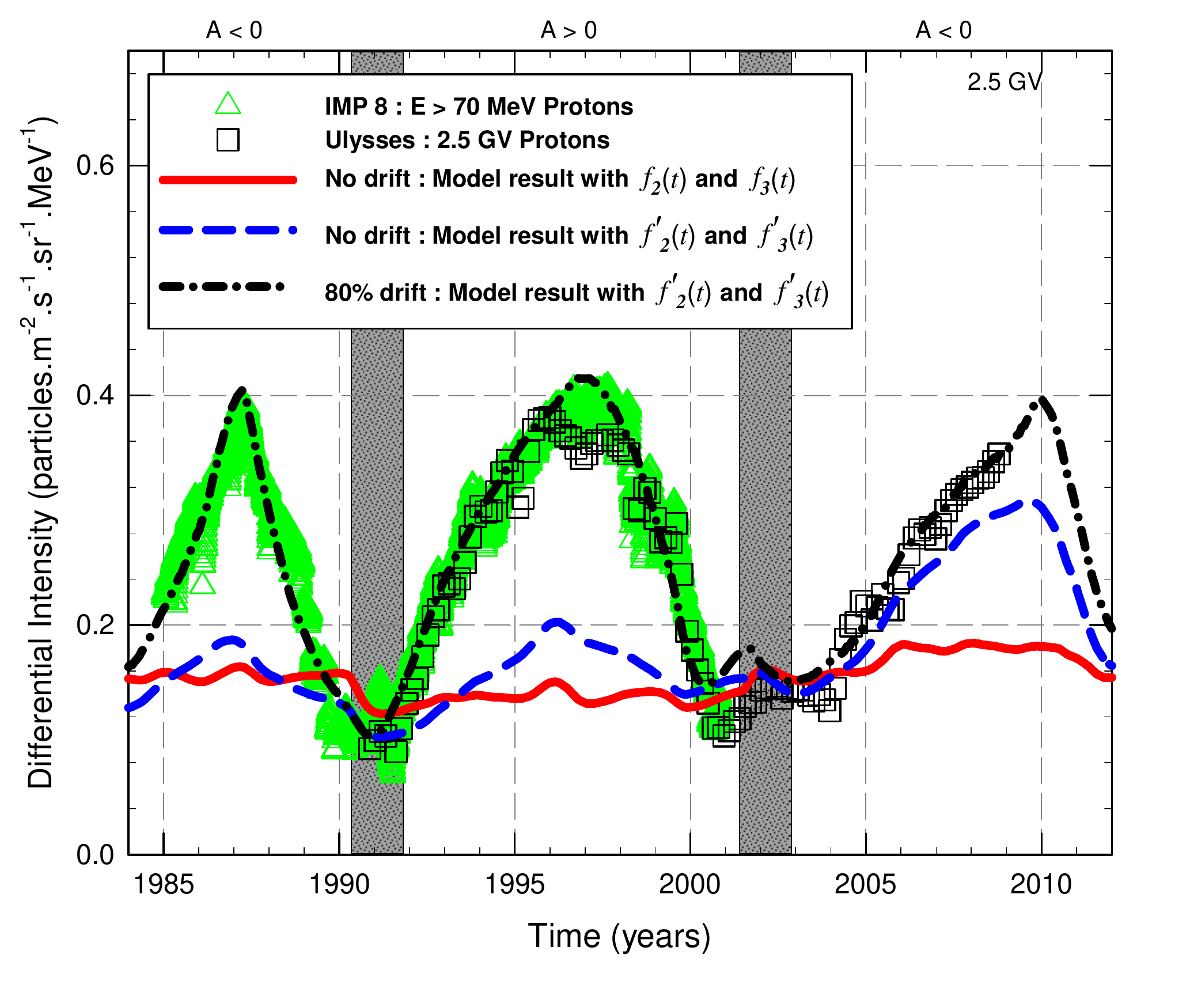}
\end{center}
\caption{Computed 2.5 GV proton intensities at Earth since 1984 are shown for two no-drift scenarios assuming theoretical time-dependent functions [$f_{2}(t)$ and $f_{3}(t)$] and modified time dependences [$f'_{2}(t)$ and $f'_{3}(t)$]. A third scenario with the latter functions and with 80\,\% drift reproducing the proton observations at Earth for $\sim$ 2.5 GV from \textit{Ulysses} (squares) \citep[][]{Heber-etal-2009} and $E > $ 70 MeV from \textit{IMP 8} (triangles) (from \protect\url{astro.nmsu.edu}).}
\label{RTF-RTHE-no-drift}
\end{figure}
 
In this section, the time dependence in diffusion coefficients is also modified by changing their rigidity dependence, inspecting Equations (\ref{E1}) and (\ref{perp}). Instead of arbitrarily choosing a different time dependence or phenomenologically constructing one by comparing computed intensities with observations, $f_{2}(t)$ and $f_{3}(t)$ are modified to reflect at 2.5 GV the dependence of Equations (\ref{E1}) and (\ref{perp}) which are applicable to higher rigidities, \textit{e.g.}  $\gtrsim 4$ GV. Due to the different rigidity dependence of the terms in Equation (\ref{E1}), which depend differently on $\delta B^2$ and $B$, the time dependence is changing as a function of rigidity.

For high rigidities, \textit{e.g.} $\gtrsim 4$ GV, the term $\frac{b_{k}}{4\sqrt{\pi}}+\frac{2}{\sqrt{\pi}(2-s)(4-s)}\frac{b_{k}}{R^s}$ in Equation (\ref{E1})  can be approximated to be a constant [$C$] \citep{Manuel-2011a,Manuel-2011b}, and one can write 
\begin{eqnarray}\label{E2}
\lambda_{||} = \frac{3s}{\sqrt{\pi}(s-1)} \frac{R^2}{b_{k} \ k_{\mathrm{min}}}\left(\frac{B}{\delta{B_{\mathrm{slab},x}}}\right)^2 C,
\end{eqnarray}
which results in a time dependence for $\lambda_{||}$ as
\begin{eqnarray}\label{parApprox2}
\lambda_{||} \propto \left( \frac{1}{\delta{B_{\mathrm{slab},x}}}\right)^2.
\end{eqnarray}
See also \cite{Manuel-2011a} and \cite{Manuel-2011b}. Note that $B$ in Equation (\ref{E2}) is cancelled by the $B$ in the expression for $R_{\mathrm{L}}$ to give Equation (\ref{parApprox2}), from which the function $f'_{2}(t)$, can be written as
\begin{eqnarray}\label{f2new}
f'_{2}(t)= C_{4}\left( \frac{1}{\delta{B}(t)}\right)^2.
\end{eqnarray}

For perpendicular diffusion, it can also be assumed for $P \gtrsim 4$ GV that
\begin{eqnarray}\label{perpApprox2}
\lambda_{\bot} \propto \left(\frac{\delta{B_{\mathrm{2D}}}}{B}\right)^\frac{4}{3}\left(\frac{1}{\delta{B_{\mathrm{slab},x}}}\right)^\frac{2}{3}.
\end{eqnarray}
From Equation (\ref{perpApprox2}), a modified function $f'_{3}(t)$, can be deduced as
\begin{eqnarray}\label{f3new}
f'_{3}(t)=C_{5}\left(\frac{\delta{B(t)}}{B(t)}\right)^\frac{4}{3}\left(\frac{1}{\delta{B}(t)}\right)^\frac{2}{3}.
\end{eqnarray}

A comparison between $f_{2}(t)$ and the new $f'_{2}(t)$ (time-dependence in parallel diffusion coefficient) and  $f_{3}(t)$ and the new $f'_{3}(t)$ (time-dependence in perpendicular diffusion coefficient) is shown in Figure \ref{RTHE-TimeDep}. The functions $f'_{2}(t)$ and $f'_{3}(t)$ show a larger difference between solar minimum and solar maximum when compared to the previous functions $f_{2}(t)$ and $f_{3}(t)$. This modified time dependence is closer to the traditional compound approach as constructed by \cite{Ferreira-and-Potgieter-2004} where the time dependence in all the transport coefficients roughly change by a factor of ten between solar minimum and maximum.

Computed intensities using  $f'_{2}(t)$ and $f'_{3}(t)$ are now compared to results from  $f_{2}(t)$ and $f_{3}(t)$ and are shown in Figure \ref{RTF-RTHE}. It is shown that there is no significant difference between the different scenarios apart from after $\sim$2004 at Earth. As shown, the introduction of  $f'_{2}(t)$ and $f'_{3}(t)$ in the model gives a better compatibility between the observations and the computed intensities after $\sim$2004 at Earth. Therefore, for this particular polarity cycle the amplitude between solar minimum and maximum intensities as presented in the various diffusion coefficients given by $f_{2}(t)$ and $f_{3}(t)$ is too small. Evidently, a larger amplitude is necessary to compute realistic modulation, as given by $f'_{2}(t)$ and $f'_{3}(t)$.

\subsection{The Effect of a Modified Time Dependence of $f'_{2}(t)$ and $f'_{3}(t)$ on Computed Intensities}

In Figure \ref{RTF-RTHE-no-drift}, two no-drift scenarios using the functions $f_{2}(t)$ and $f_{3}(t)$ are compared to using the modified functions  $f'_{2}(t)$ and $f'_{3}(t)$. A further scenario, with $K_{A0}=0.8$ using the modified functions  $f'_{2}(t)$ and $f'_{3}(t)$, is also shown at Earth from 1984 onwards. In comparison the $\sim$2.5 GV \textit{Ulysses} and $E > $ 70 MeV \textit{IMP 8} proton observations are shown. It follows that the no-drift $f'_{2}(t)$ and $f'_{3}(t)$ scenarios give a computed intensity amplitude between solar minimum and maximum which is much larger, especially after $\sim$2004 onwards, compared to the previous assumptions using $f_{2}(t)$ and $f_{3}(t)$. Using the modified functions $f'_{2}(t)$ and $f'_{3}(t)$ with $K_{A0}=0.8$ gives a compatible result at Earth from $\sim$2004 onwards, illustrating that a larger time dependence in the magnitude of the diffusion coefficients are needed over this particular solar cycle. This means that for this particular solar cycle at Earth, time-dependent changes in the diffusion coefficients are more important compared to previous cycles. This can be seen by first comparing the dashed blue line with the solid red line showing a much larger modulation amplitude, and then comparing the dashed--dotted black line, to previous attempts. For this particular solar minimum, the drift effects are downplayed by changes in the diffusion coefficients especially through a time dependence in their rigidity dependence. This aspect of the recent solar minimum period was also discussed in detail by \cite{Potgieter-etal-2013}.

\section{Summary and Conclusions}
A well established time-dependent cosmic-ray modulation model was further utilised to compute long-term modulation over multiple consecutive solar activity and magnetic cycles. Results were compared to \textit{Voyager 1}, \textit{Ulysses}, and \textit{IMP 8} proton observations to establish whether the parameters assumed in this work result in realistic computed intensities. The current sheet tilt angle values and magnetic field magnitude measurements at Earth were used as input for the model. Also a statistical variance in the magnetic field was calculated and together with the measured magnetic field used to construct a time dependence in the transport parameters as given by the theoretical studies of \citet{Teufel-and-Schlickeiser-2002,Teufel-and-Schlickeiser-2003}, \citet{Shalchi-etal-2004a}, and \citet{Minnie-etal-2007b}. This time-dependent change in the various modulation parameters is then transported out into the simulated heliosphere resulting in computed intensities compatible to observed cosmic-ray intensities on a global scale. 

This new approach successfully produces cosmic-ray intensities along the \textit{Voyager 1} trajectory that are compatible to observations until 2012 and also gives intensities at Earth that are compatible to the \textit{IMP 8} and \textit{Ulysses} observations until $\sim$2004. However, after $\sim$2004, at Earth, the computed intensities failed to reproduce the observations, giving lower intensities than observed. A thorough parameter study was subsequently conducted by testing the effects of different modulation parameters such as the heliopause position, the TS position, and in particular different time-dependent diffusion coefficients etc., on the computed intensities. It was shown that solar-cycle related changes in these parameters did not lead to improved compatibility with the data for the period after $\sim$2004 at Earth. 

These calculations suggest modifications to the assumed time dependence in the diffusion coefficients. It was also found that variations in the HPS compared to what is assumed in this work can always be compensated for by an increased heliopause position and/or by changing the diffusion coefficients.  The effect of the drift coefficient on cosmic-ray modulation was also investigated and a modification to the time-dependent function, which scales drifts over a solar cycle, was proposed. Although this modified function allows a faster recovery of drifts towards solar minimum, it was not sufficient to compute compatible intensities after $\sim$2004.

The time dependence in the parallel and perpendicular diffusion coefficients was consequently modified. This led to compatible computed intensities along the \textit{Voyager 1} trajectory and at Earth even for the period after $\sim$2004. Cosmic-ray modulation especially for the present polarity cycle and the recent solar minimum period is no longer largely determined by changes in only the drift coefficient and the tilt angle of the current sheet but also by changes in the diffusion coefficients over time.

\begin{acks}
This work is partially supported by the South African National Research Foundation (NRF). The authors wish to thank du Toit Strauss and Eugene Engelbrecht for valuable discussions, and the editor and reviewer for the constructive comments. We used data from the Wilcox Solar Observatory (\url{wso.stanford.edu}), NASA COHOWeb (\url{cohoweb.gsfc.nasa.gov}), W. R. Webber (\url{astro.nmsu.edu}), and \cite{Heber-etal-2009}.
\end{acks}


\end{article}
\end{document}